\documentclass[a4paper, 12pt]{article}
\pdfoutput=1

\usepackage{amsmath}
\usepackage{amsfonts}
\usepackage{amssymb}
\usepackage[final]{graphicx}
\usepackage{graphics}
\usepackage{color}
\usepackage{hyperref}       
\usepackage{url}            

\graphicspath{ {./figs/} }
\usepackage{comment}
\usepackage{multicol}
\usepackage{multirow}
\usepackage{datetime}
\usepackage{anysize}
\usepackage[toc, page,titletoc]{appendix}
\usepackage{chngcntr}

\definecolor{bluscuro}{rgb}{0.15, 0.2, .85}
\hypersetup{final, colorlinks, citecolor=bluscuro, linkcolor= bluscuro, urlcolor=bluscuro}

\marginsize{2cm}{2cm}{2cm}{2cm}
\def\be   {\begin{equation}}   
\def\ee   {\end{equation}}
\def\nn   {\nonumber}

\begin{document}

\pagestyle{empty}

\vspace{0.5cm}
\begin{center}
{\LARGE \textbf{
The impact of undetected cases on \\
tracking epidemics: the case of COVID-19
}\par}

\vspace{1.5cm}
{\large{\textsc{
Andrea De Simone$^{\,a,\, b,\,c,\,}$\footnote{
\href{mailto:andrea.desimone@sissa.it}{andrea.desimone@sissa.it}}
}}}, 
{\large{\textsc{
Marco Piangerelli$^{\,c,\,}$\footnote{
\href{mailto:marco.piangerelli@unicam.it}{marco.piangerelli@unicam.it}}
}}}
\\[1cm]

\large{\textit{
$^{a}$ Physics Area, SISSA, Trieste, Italy \\  
$^{b}$ INFN, Sezione di Trieste, Italy \\  
$^{c}$ School of Science and Technology,
  University of Camerino, Italy
}}
\end{center}

\vspace{0.5cm}

\begin{center}
\textbf{Abstract}
\begin{quote}
One of the key indicators used in tracking the evolution of an infectious disease is the reproduction number. 
This quantity is usually computed using the reported number of cases, but ignoring that many more individuals may be infected (e.g. asymptomatics).
We propose a statistical procedure to quantify the 
impact of undetected infectious cases on the determination of
the effective reproduction number.
Our approach is stochastic, data-driven and not relying on any compartmental model.
It is applied to the COVID-19 case in eight different countries and all Italian regions, showing that the effect of undetected cases leads to estimates of the effective reproduction numbers larger than those obtained only with the reported cases by factors ranging from two to ten.
Our findings urge caution about deciding when and how to relax containment measures based on the value of the reproduction number.
\end{quote}
\end{center}

\newpage
\setcounter{page}{1}
\pagestyle{plain}
\tableofcontents

\section{Introduction}

Tracking the evolution of the spread of an infectious disease is of primary importance during the whole course of any epidemic.
An accurate evaluation of the transmission potential of the disease  
provides invaluable information to guide the decision-making process of control interventions, and to assess their effectiveness.
One of the key epidemiological variables in this respect is
the effective reproduction number $R$, defined as 
the average number of secondary cases per primary case of infection.
In order to stop an epidemic, $R$ needs to be persistently reduced to a level below 1.
The issue of providing reliable estimates of $R$ is particularly severe now during the on-going COVID-19 pandemic \cite{Who}, and urgently calls for efforts towards a comprehensive mathematical modelling of the outbreak.

In this paper we propose a statistical framework for computing the effective reproduction number $R$ characterized by the following main features.
\begin{enumerate}

\item \textit{Stochastic.} 
Our approach is stochastic, and not rooted in any deterministic framework of compartmental models, such as SIR and its extensions.
Although compartmental  models can indeed provide very useful outcomes,
especially for extrapolating the outbreak evolution into the near future, they rely on the simultaneous determination of all the coefficients appearing in the differential equations describing the dynamics of each compartment. 

\item \textit{Real-time.} The method provides a time series of estimations of $R$ at each time step (e.g. one day), and not a single \textit{a posteriori} value when the outbreak is almost over.

\item \textit{Bayesian.}
Within a Bayesian framework the results have a transparent probabilistic interpretation, the assumptions (priors) are explicit and their role is clearly tracked.
A Bayesian updating procedure also accounts for the real-time evolution of the probabilities.

\item \textit{Comprehensive.} 
Our method is explicitly carried out for taking into account the (unknown) number of undetected cases. However, it can be straightforwardly generalized to include
any additional random variable affecting the reproduction number.
Since the reproduction number is a highly complex quantity, affected by a wide variety of factors, such as biological, environmental and social factors \cite{delamater2019complexity}, this feature is particularly relevant.
\end{enumerate}
Several methods  for estimating the reproduction number have been proposed in the literature and have one or more of the above characteristics
(see e.g. Refs.~\cite{Teunis2004, bettencourt,Cori2013, Cori2019, Li489,bardina2020stochastic,Giordano_2020}).
But, to the best of our knowledge, our approach is the first one combining  all those components at once.

In particular, in this paper we are interested in assessing the role and impact of the number of undetected infection cases onto the effective reproduction number. 
There may be different reasons why an infected patient is undetected and does not appear in the official reports: 
individuals not showing the symptoms of the disease but are able to infect others (asymptomatics), individuals whose symptoms have not been linked to the disease under consideration (especially in the early stages of the outbreak), 
 impossibility of a complete population screening, etc.

For COVID-19, the number of undetected cases of infection may indeed be rather large.
 According to Ref.~\cite{Lavezzo2020_Vo}, in the small Italian town of Vo' Euganeo 43\% of the confirmed infections were asymptomatic with no statistically different  viral load with respect to the symptomatic cases.
Another study performed at the New York–Presbyterian Allen Hospital and Columbia University Irving Medical Center in NYC, pointed out that 29 of the 33 patients who were positive for SARS-CoV-2 (87.9\%) had no COVID-19 symptoms~\cite{NEJMc2009316}. 
These results  justify the denomination of asymptomatic transmission of SARS-CoV-2 as the ``Achilles' heel'' of current COVID-19 containment strategies \cite{NEJMe2009758}.

\section{Methodology and Results}
\label{sec:methods}

We develop a Bayesian statistical formulation of the evolution of the effective reproduction number, 
building upon the works of Refs.~\cite{bettencourt, Cori2013, Cori2019}.
Full details about our model and calculations are provided in Supplementary Material \ref{app:statmodel}.
It is important to remark that our approach is stochastic and data-driven, not relying on any deterministic compartmental model.

The number of new infected individuals at a given time are modelled as a discrete-time stochastic process.
Among the many variables affecting the effective reproduction number $R$ \cite{delamater2019complexity}, we choose as the most relevant ones
 the time series of disease incidence data up to time $t$
($I_{\leq t}=\{I_0,I_1,\ldots, I_t\}$), and the serial interval ($W$), i.e.
the time from symptom onset in a primary case to symptom onset of his/her secondary cases.

\begin{figure}[t] 
    \centering
    \includegraphics[width=0.95\textwidth]{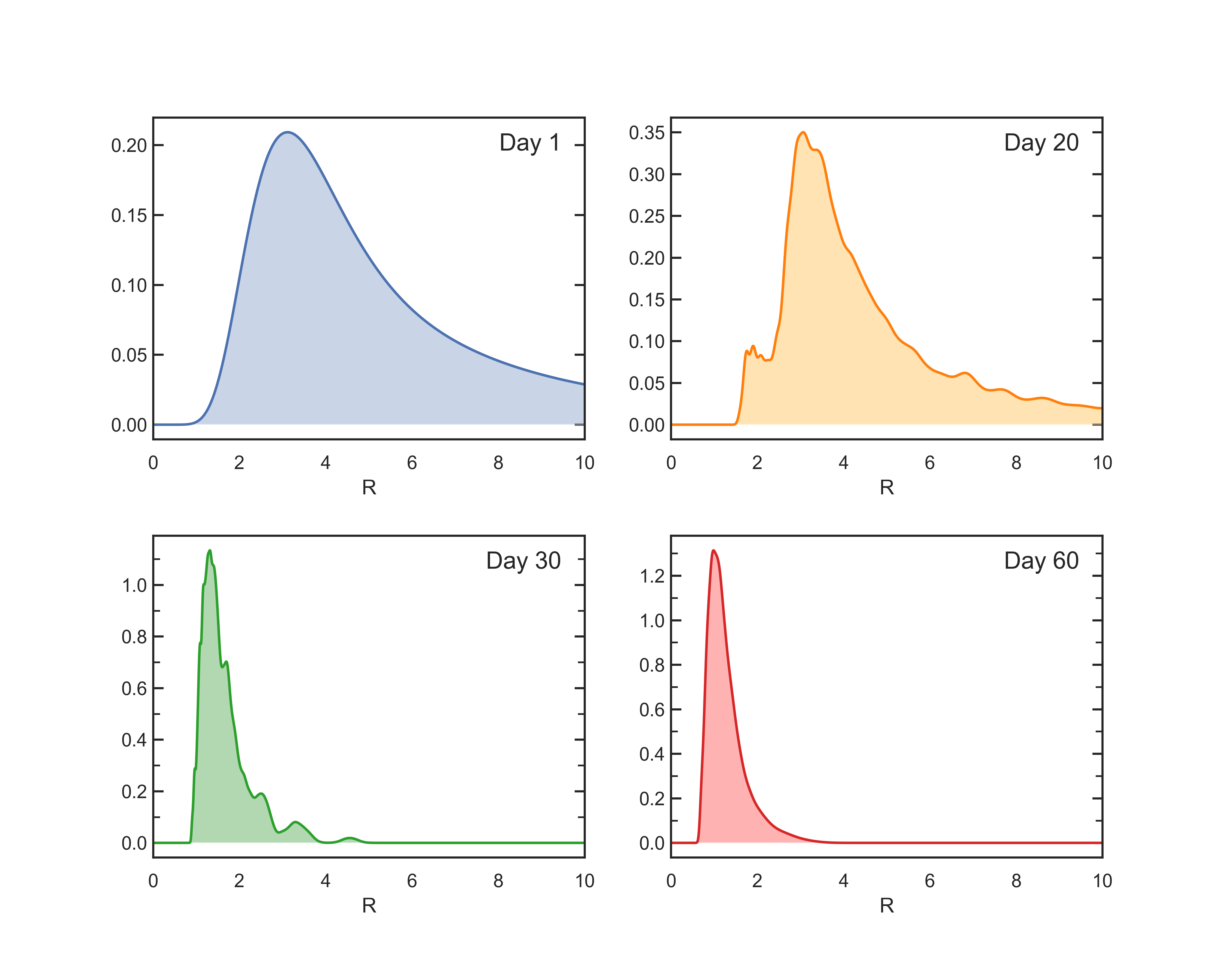}
\caption{
{\small \emph{Time evolution of the posterior probability density of the effective reproduction number $R$ (we used the South Korea data, just for illustration purpose). The day 0 corresponding to a uniform prior is not shown.
}}}
\label{fig:posteriorevolution}
\end{figure}

At any time $t$ during an epidemic, the posterior probability density of $R$, conditioned on the past incidence data and serial interval, $p_{R|I_{\leq t}, W}(r|i_{\leq t}, w)$ 
encodes a great deal of information about the current state of the outbreak.
In particular, the effective reproduction number  at time $t$ ($R_t$)
can be derived from it as the expected value 
\be 
R_t=\textrm{E}[R|I_{\leq t},W]=\int dr\, r\, p_{R|I_{\leq t}, W}
(r|i_{\leq t}, w)\,,
\label{eq:Rtexpected}
\ee 
as well as the  95\% central credible intervals (see Fig.~\ref{fig:posteriorevolution}). 
As initial prior for $R$  we adopt an uninformative uniform distribution throughout the paper.

\begin{figure}[t!] 
    \centering
    \includegraphics[width=0.5\textwidth]{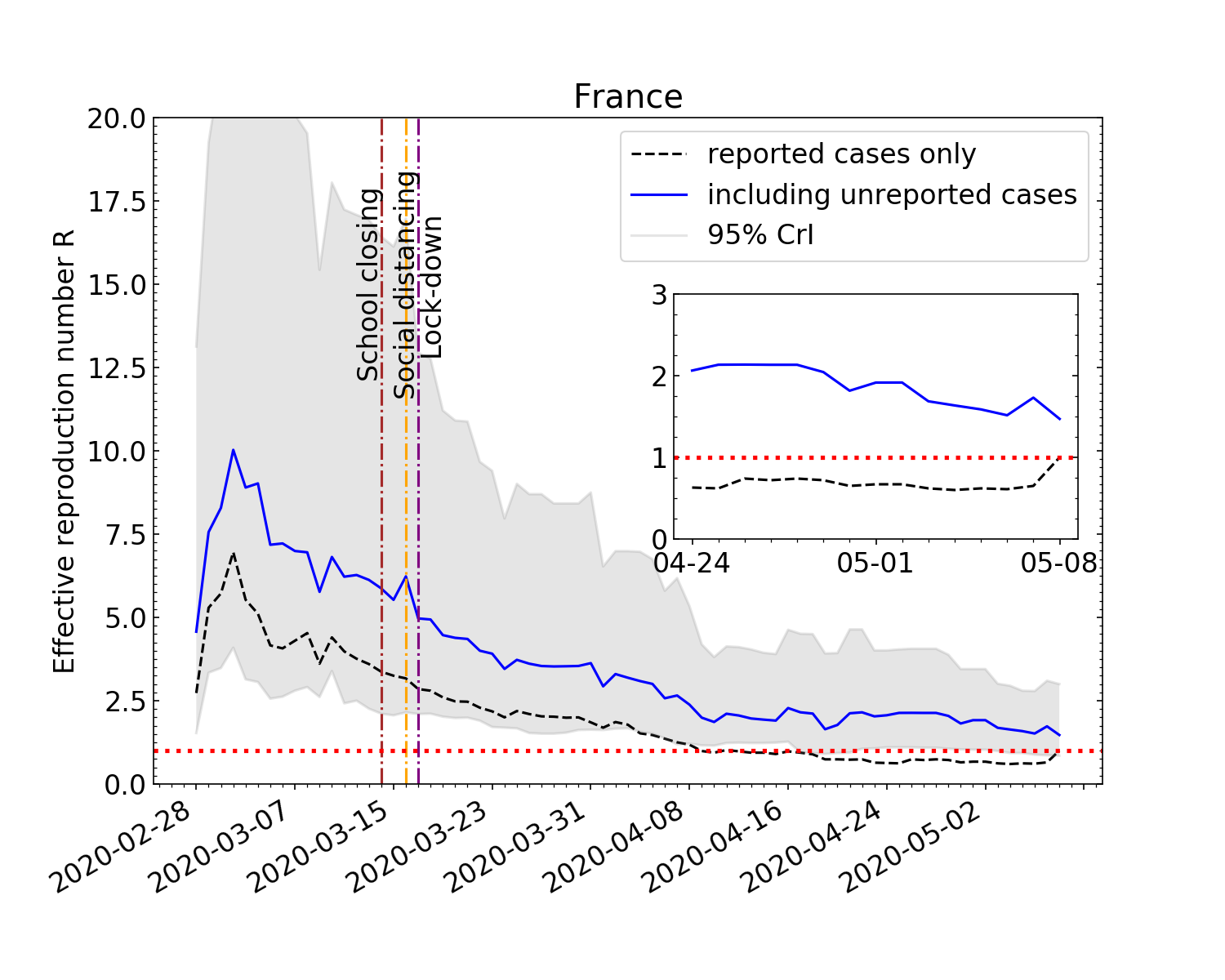}
    \hspace{-0.5cm}
    \includegraphics[width=0.5\textwidth]{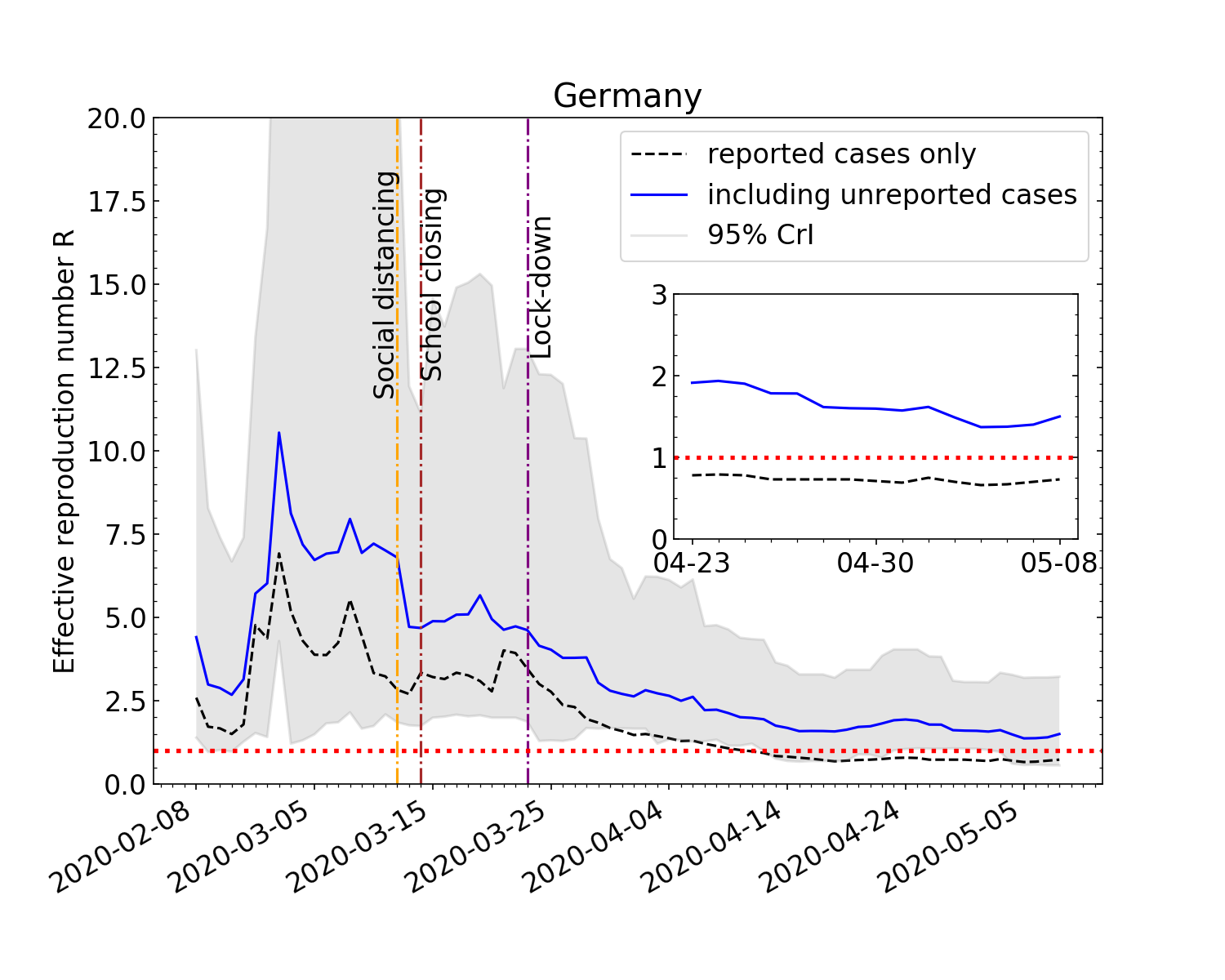}\\
    \includegraphics[width=0.5\textwidth]{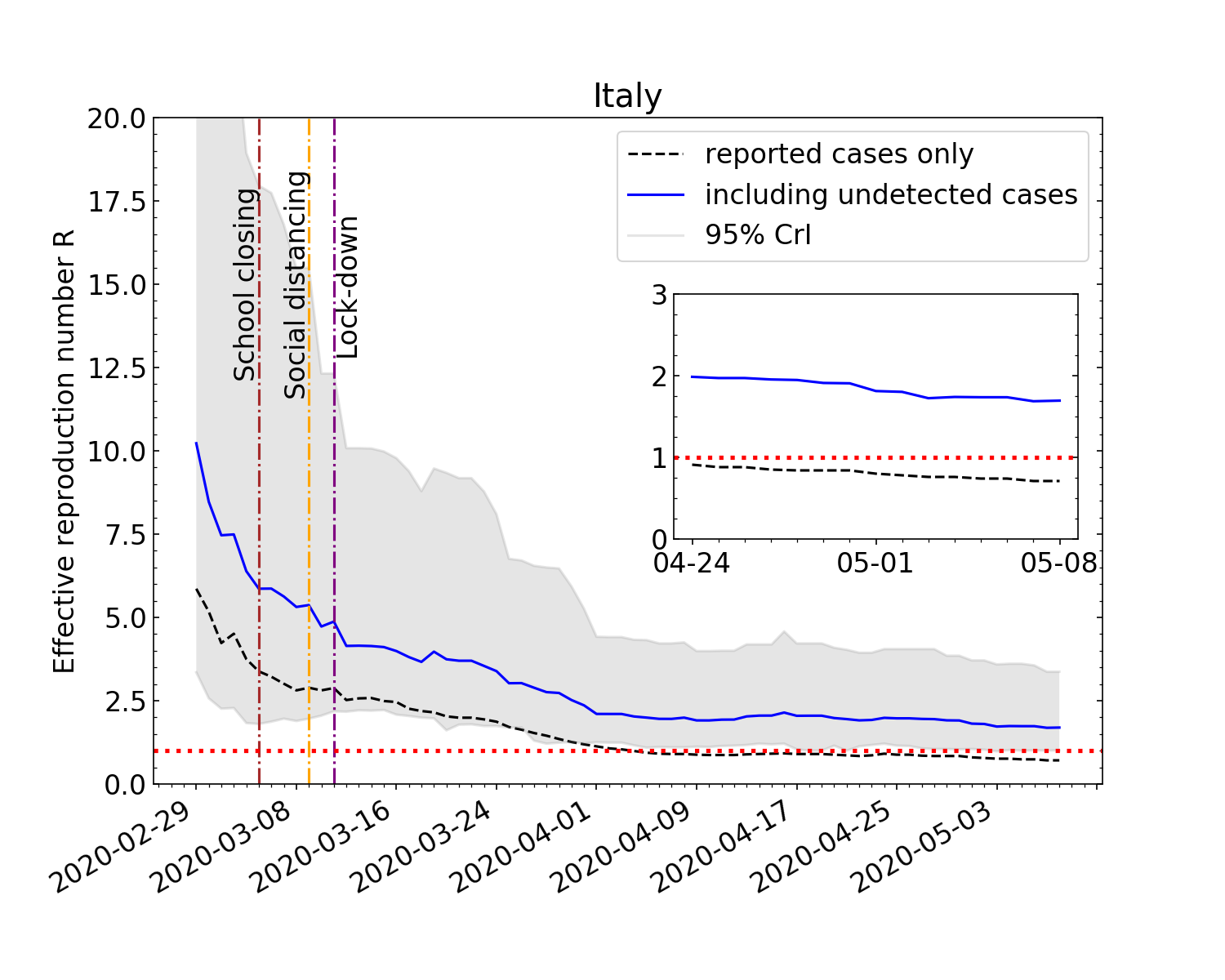}
    \hspace{-0.5cm}
    \includegraphics[width=0.5\textwidth]{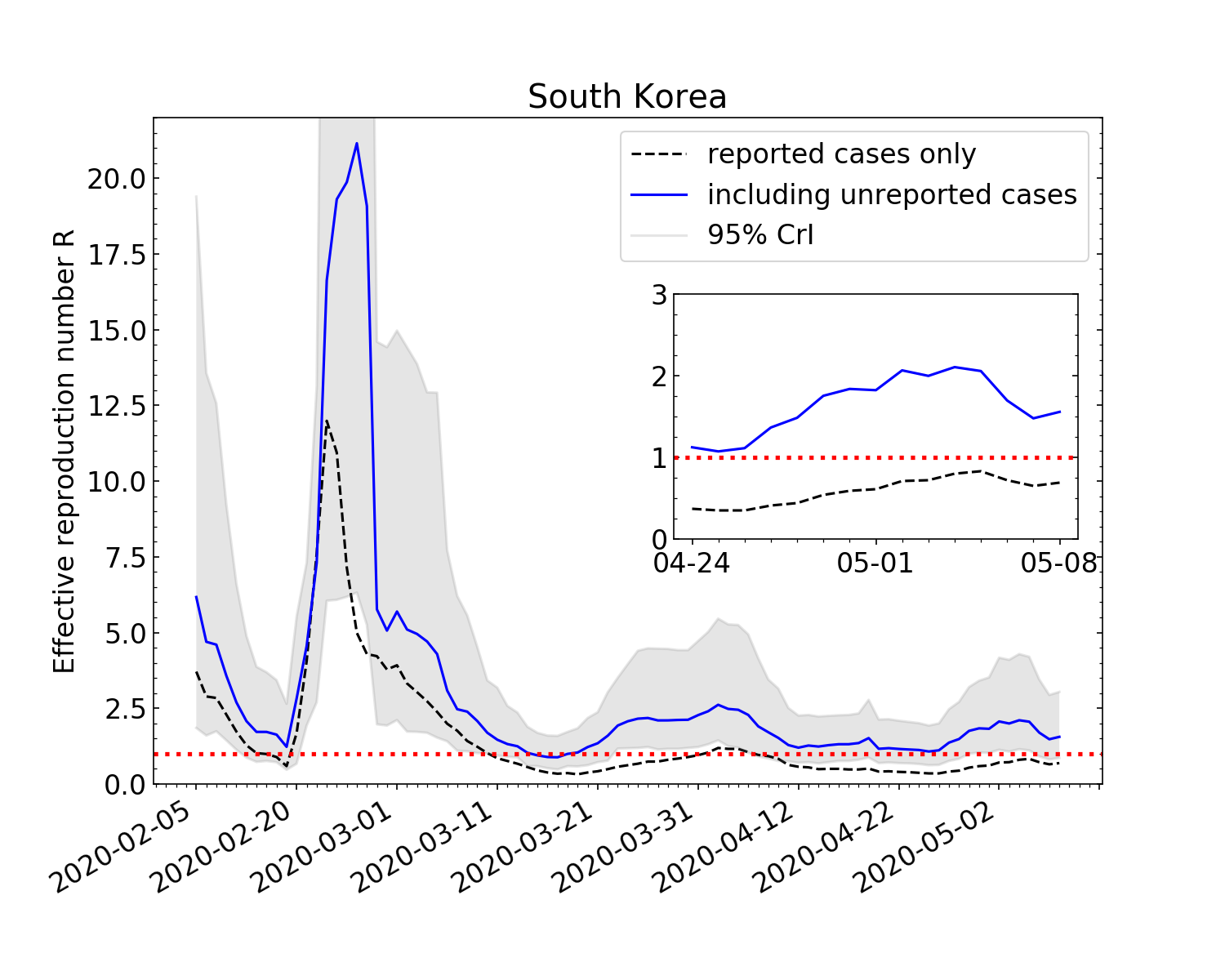}
\caption{
{\small \emph{Time evolution of effective reproduction number for COVID-19
in France, Germany, Italy and South Korea. 
\emph{Black dashed line:} result of including only the reported cases of infection.
\emph{Blue solid line:}  mean of the posterior probability marginalized over the undetected cases and the serial interval parameters.
\emph{Gray shaded area:} 95\% central credible interval.
The inset shows the results of the past two weeks in greater detail.
The vertical lines refer to the time when containment measures have been adopted.
}}}
\label{fig:countries1}
\end{figure}

Our mathematical formalism is general and flexible enough to incorporate in a straightforward way 
any other  random variable affecting $R$, in addition to the incidence and serial interval usually considered. In particular, in this paper we  consider the impact on $R$ due to 
the unknown number of undetected cases, provided we make assumptions about  their probability distribution. The general posterior probability density of $R$ needed in Eq.~(\ref{eq:Rtexpected}),  given Poisson-distributed incidence data, serial interval data and the undetected cases is reported in Eq.~(\ref{eq:postR3}), including the marginalization over all nuisance parameters.
Our results allow one to track the effective reproduction number in real time during an outbreak, including the  effects of undetected cases, under general assumptions.

We also took a further step by assuming a parametric form for the serial interval distribution, and uniform probabilities for the undetected cases. 
In this simple case we were able to compute the posterior density $p_{R|I_{\leq t}, W}(r|i_{\leq t}, w)$ analytically in close form (see Eq.~(\ref{eq:postR4})). 
This is the form we are applying to the COVID-19 data, as we now turn to discuss.

\begin{figure}[t!] 
    \centering
    \includegraphics[width=0.5\textwidth]{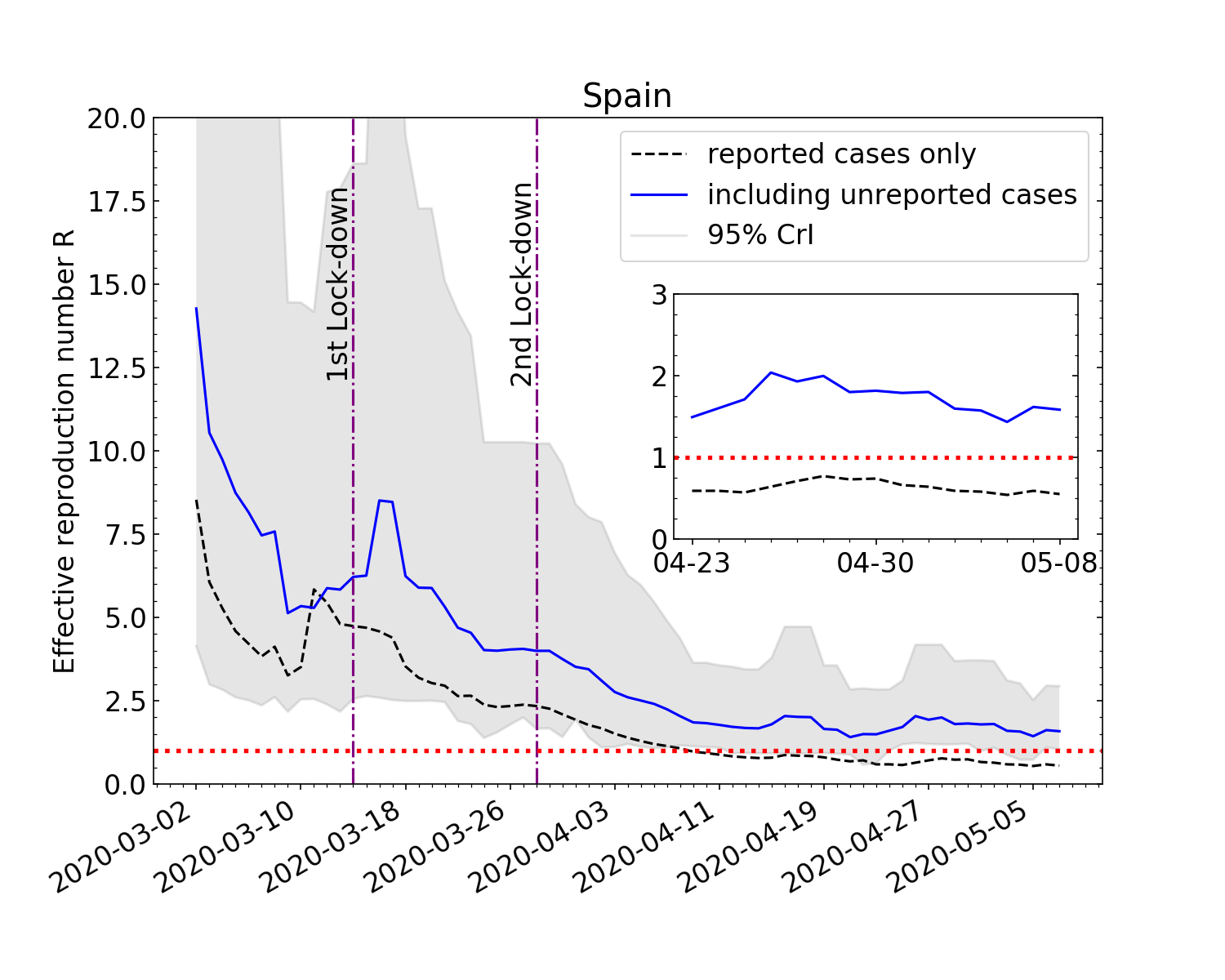}
    \hspace{-0.5cm}
    \includegraphics[width=0.5\textwidth]{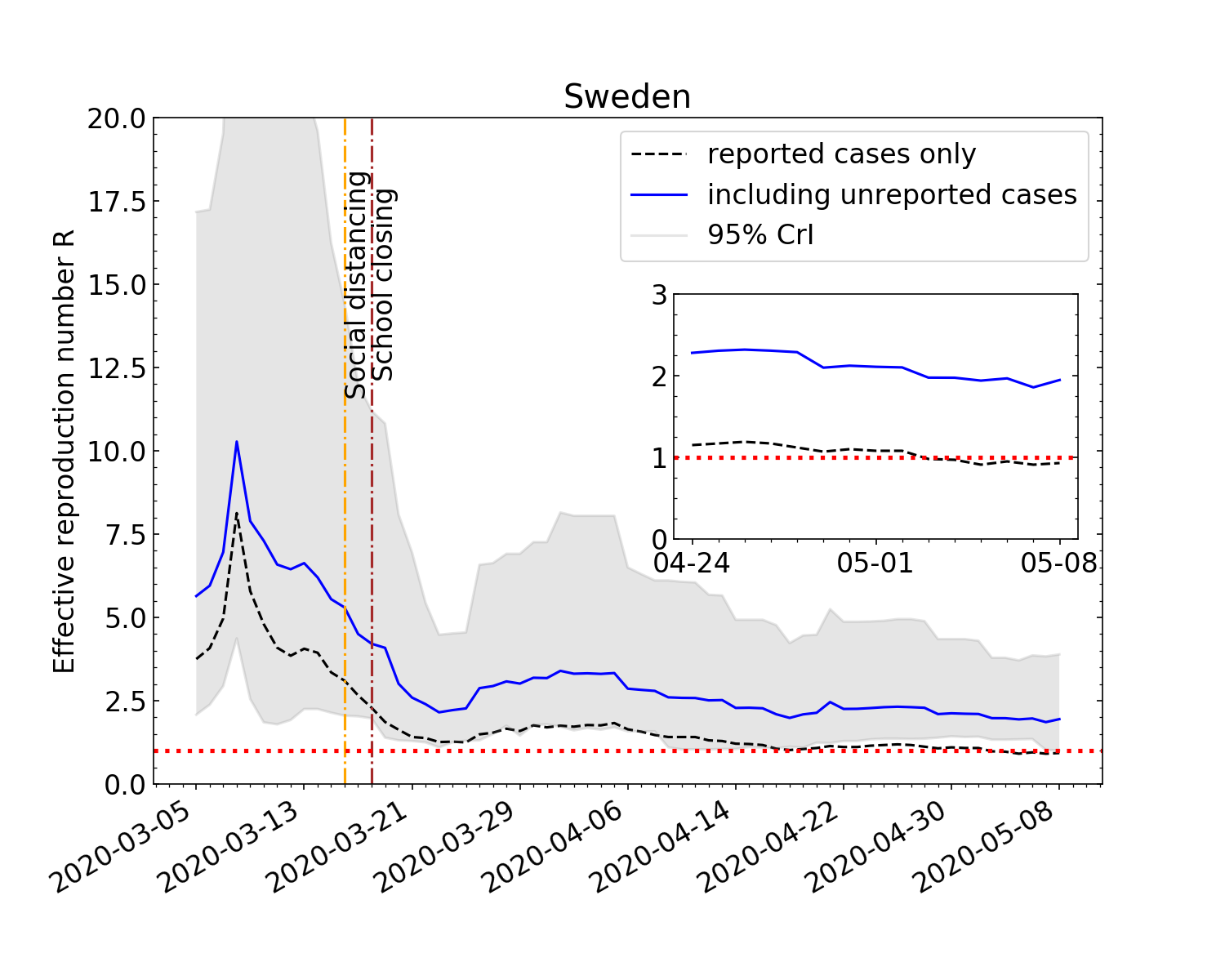}\\
    \includegraphics[width=0.5\textwidth]{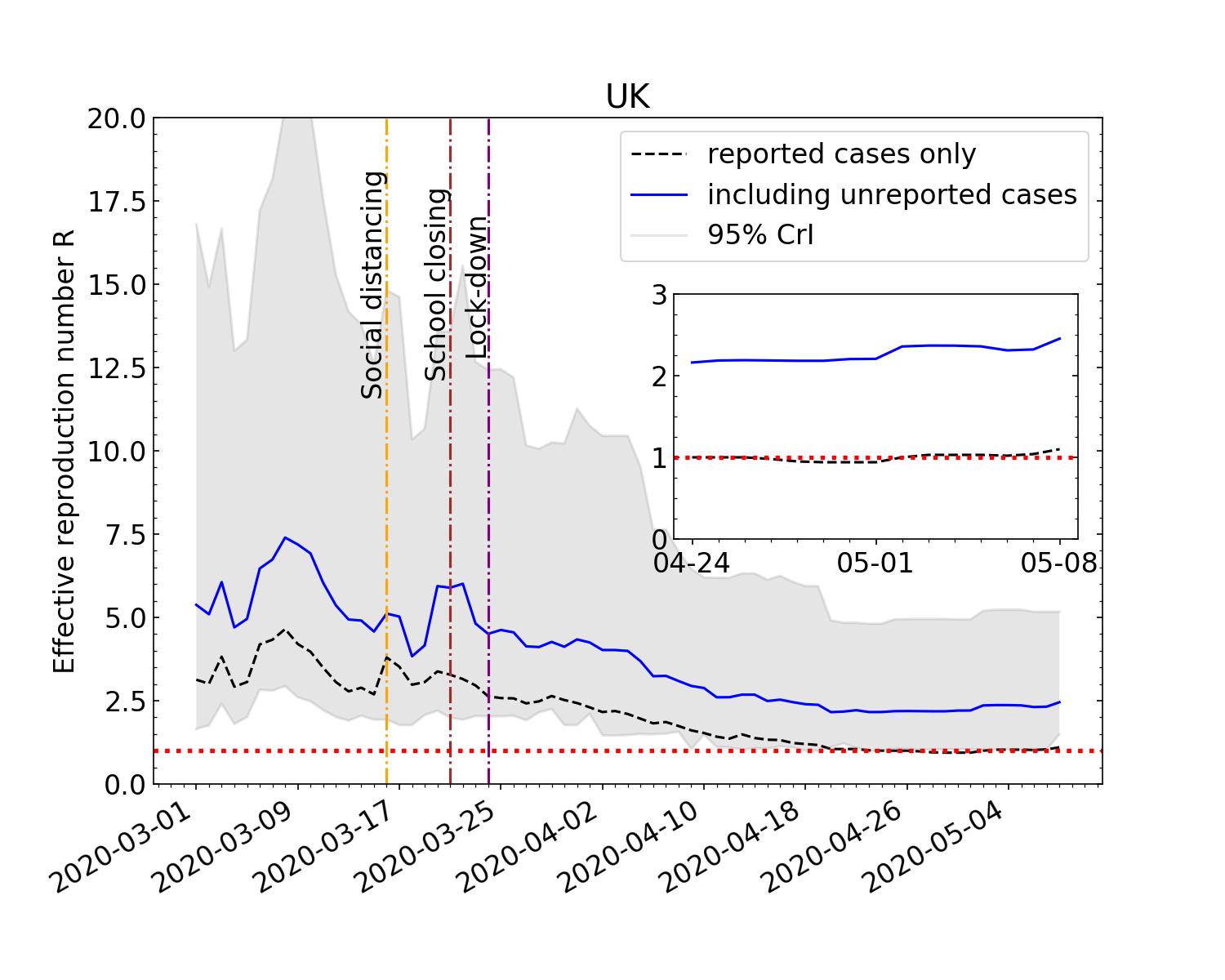}
    \hspace{-0.5cm}
    \includegraphics[width=0.5\textwidth]{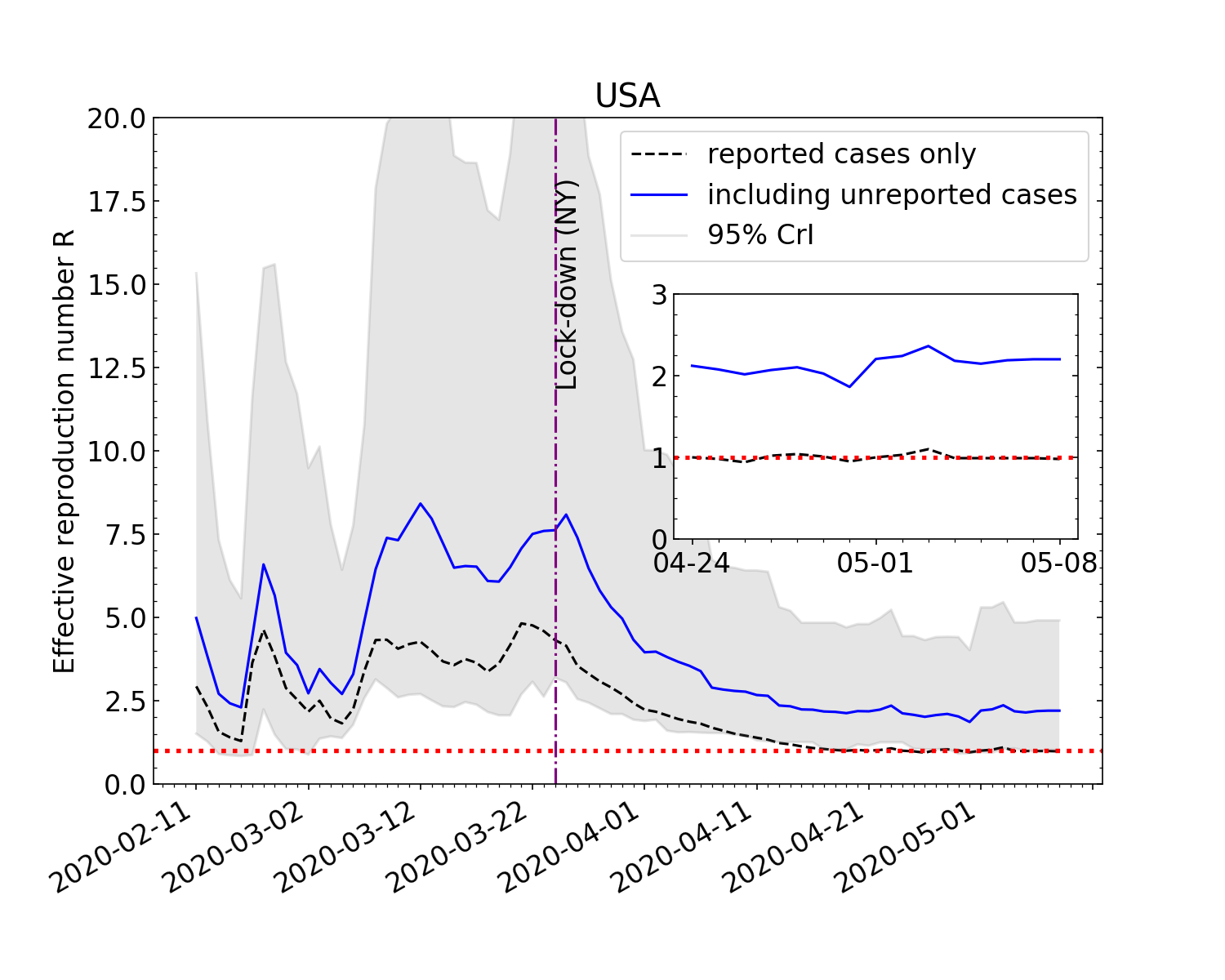}
\caption{
{\small \emph{Time evolution of effective reproduction number for COVID-19
in Spain, Sweden, UK and USA. 
\emph{Black dashed line:} result of including only the reported cases of infection.
\emph{Blue solid line:}  mean of the posterior probability marginalized over the undetected cases and the serial interval parameters.
\emph{Gray shaded area:} 95\% central credible interval.
The inset shows the results of the past two weeks in greater detail.
The vertical lines refer to the time when containment measures have been adopted (for the USA,
we show the lock-down date in the state of NY as a reference).
}}}
\label{fig:countries2}
\end{figure}

We employ COVID-19 daily incidence data
for eight different countries (France, Germany, 
 Italy, South Korea, Spain, Sweden, UK and USA), 
retrieved from Ref.~\cite{ECDC_eu}.
In Supplementary Material \ref{app:regions} we also analyze
the incidence data in the twenty Italian regions.

For the serial interval distribution, we adopt a gamma distribution 
with the normally-distributed parameters estimated in Ref.~\cite{lombardy} (see Eqs.~(\ref{eq:gammadistrib}) and (\ref{eq:gammadistrib1})),
which is also consistent with other determinations in several countries (see Ref.~\cite{park2020systematic}).
We assume a discrete uniform distribution for the number of undetected cases, allowing them to be  at most a factor of $C_U$ times the number of reported cases at each time. The marginalization of the posterior probability of $R$ over $C_U$ is cutoff at $C_U=2$. Of course this choice of the prior for $C_U$ is subjective, and we believe it is also rather conservative.

The computation of $R$ starts from the first day reported on the dataset. It is worth mentioning that the starting date is not the same for all the countries taken into account.
The results are shown in Fig.~\ref{fig:countries1} and Fig.~\ref{fig:countries2}. The figures show the time evolution of $R$ and, wherever applicable, the dates when contagion containment measures have been enforced.
In Table \ref{tab:results} we report the numerical results for the last day of our analysis.

Our results clearly show that $R$, after a transitory period, is in a down trend in all the countries we considered.
In all the countries considered, we find values of the mean value 
$R_t$ larger than the ones without considering the undetected cases by factors of about 2 -- 4. This is a somewhat expected consequence of our conservative choice of $C_U$ being at most 2. 
By allowing more undetected cases, the resulting reproduction numbers would be necessarily higher.
Furthermore, the upper values of the 95\% credible intervals 
are larger than the estimate with only officially reported cases  by factors up to 10, and they are all significantly above 1.

\begin{table}[t]
\centering
\begin{tabular}{|c|c|c|c|}
\hline
\multirow{2}{*}{\textbf{Country}} & 
$R$ \textbf{with reported} &
\textbf{Mean of  marginalized  } &
\multirow{2}{*}{\textbf{95\% CrI}} \\
& \textbf{cases only} & 
\textbf{posterior probability} & \\
\hline 
France & 1.00 & 1.47 & (0.85,\, 3.00)\\
\hline
Germany & 0.73 & 1.50 & (0.56,\, 3.22)\\
\hline
Italy & 0.71 & 1.69 & (1.01,\, 3.37)\\
\hline
South Korea & 0.69 & 1.56 & (0.86,\, 3.04)\\
\hline
Spain & 0.55 & 1.58 & (1.04,\, 2.94)\\
\hline
Sweden & 0.93 & 1.95 & (1.01,\, 3.89)\\
\hline
UK & 1.10 & 2.45 & (1.49,\, 5.17)\\
\hline
USA & 0.98 & 2.20 & (1.02,\, 4.91)\\
\hline
\end{tabular}
\caption{
{\small \emph{
The values of the effective reproduction number $R$ for COVID-19, on  the last day of our analysis {2020-05-08), for each of the countries we considered.
The second column reports the value we find by including only reported incidence data and mean values of the serial interval distribution. On the third column we report the mean of the posterior distribution of $R$, marginalized over the nuisance parameters describing the serial interval and undetected cases distributions.
The corresponding 95\% credible interval is reported on the last column.}
}}}
\label{tab:results}
\end{table}

\section{Conclusions and Outlook}

In this paper we took the first steps towards a comprehensive stochastic modelling of the effective reproduction number $R$ during an epidemic.
We followed a completely general approach, which 
enables one to  account for any random variable affecting $R$.
In particular, our primary focus was on assessing the impact on $R$ of the number of undetected infection cases.

We investigated the time evolution of the posterior probability density of $R$, marginalized over the parameters of the serial interval  and undetected cases distributions.
The application of our method to the COVID-19 outbreak in different countries show that the effective reproduction number is largely affected by the undetected cases, and in general it increases by factors of order 2 to 10.

There are several directions in which further research can be carried out, aiming at expanding the capabilities of the basic framework described in this paper.
For instance, it is desirable to explore a more realistic model of the probability distribution of undetected cases, also including time dependence.
Furthermore, it is also possible to adopt a fully data-driven approach by performing non-parametric estimation of the serial interval distribution from transmission chains data.

The stochastic approach outlined in this paper is not designed to establish or predict any cause-effect relationship between the $R_t$ trend
and  the enforcement of the containment measures.
It is nevertheless possible to use our results (especially the credible intervals)  to define some robust criterion for evaluating the effectiveness of the containment measures. 

Our findings can be used by public health institutions to take more informed
decisions about designing and gauging the strategies of infection containment.
According to our results, we recommend  caution in deciding 
when and how to relax the containment measures based on the value of the reproduction number, since it may be much larger than usually estimated.

\paragraph*{Acknowledgements}
M.P. thanks R.~De Leone for useful conversations.

\begin{multicols}{2}
\begin{footnotesize}
\bibliographystyle{ieeetr}  
\bibliography{references}  
\end{footnotesize}
\end{multicols}

\appendixtitleoff
\appendixpageoff
\renewcommand{\appendixtocname}{Supplementary Material}

\begin{appendices}
\setcounter{section}{19}
\setcounter{equation}{0}
\numberwithin{equation}{section}
\setcounter{figure}{0}
\counterwithin{figure}{section}
\setcounter{table}{0}
\counterwithin{table}{section}

\section*{Supplementary Material}

\subsection{Statistical model}
\label{app:statmodel}

We work in a Bayesian framework in which we treat all observations and parameters as random variables. The parameter of primary interest in this paper is the effective reproduction number $R$, considered as a continuous random variable with prior probability density function $p_R(r)$. 
We then consider the  observed incidence cases (number of new infected individuals at time $t$) $I_t$ and  the unknown number of undetected cases $U_t$ as positive integer-valued stochastic processes with discrete time index $t$.

We will also consider the stochastic process
$T_t\equiv I_t+U_t$, describing the total number of incident cases as a function of time, i.e. the sum of observed (reported) cases and the  number of undetected cases. We ignore the imported cases. Notice that, in general, $I_t$ and $U_t$ are  dependent, and their dependence is encoded by the conditional variable $U_t|I_t$.

The serial interval is described by the discrete random variable $W$.
Its probability mass function $p_W(w)$  provides the probability of a secondary case arising $w$ time steps after a primary case.

Given a time window of $\tau$ time steps, over which $R$ is assumed to be constant, we can split the times into two intervals:  $0\leq k\leq t-\tau-1$
and $t-\tau\leq k\leq t$. To avoid notational clutter, we will indicate by
$I_{< s}$ the set of random variables before time $s$
\be
I_{< s}\equiv \{I_0,I_1,\ldots I_{s-1}\}\,,
\ee
and by $I_{[t-\tau,t]}$ the set of random variables 
in the time interval $[t-\tau,t]$
\be 
I_{[t-\tau,t]}\equiv \{I_{t-\tau},\ldots, I_t\}\,.
\ee
In our numerical simulations leading to the results
displayed Sections \ref{sec:methods} and \ref{app:regions} we set $\tau=7$
days.

By Bayes' theorem, at any given time $t>\tau$, the posterior probability density of $R$ given
the serial interval distribution, the incidence data history and the undetected cases in the time window $[t-\tau,t]$ is
\begin{align}
&p_{R|I_{<t-\tau}, T_{[t-\tau,t]}, W}
(r|i_{<t-\tau},(i+u)_{[t-\tau,t]}, w) \nn\\
&\propto p_R(r)\cdot p_{I_{<t-\tau}, T_{[t-\tau,t]}, W|R}
(i_{<t-\tau},(i+u)_{[t-\tau,t]}, w|r)
\nn\\
&\propto p_R(r)\cdot p_{W|R}(w|r) \cdot 
p_{T_{[t-\tau,t]}|I_{<t-\tau},W,R}
((i+u)_{[t-\tau,t]}|i_{<t-\tau}, w,r)\nn\\
&\propto p_R(r)\cdot p_W(w)\cdot 
\prod_{k=t-\tau}^t p_{T_k|I_{<t-\tau},W,R}(i_k + u_k|i_{<t-\tau}, w,r)\nn\\
&= p_R(r)\cdot p_W(w)\cdot 
\prod_{k=t-\tau}^t p_{T_k|I_{< k},W,R}(i_k + u_k|i_{< k}, w,r)\,,
\end{align}
where we assumed that $W$ and $R$ and independent (a generic dependence between them can be implemented in a straightforward way).
Now, we can use the total law of probability to sum over the unknown number of undetected cases, 
assuming that $U_k$ depends only on $I_k$.
This way, we can get the posterior probability density for $R$
given the serial interval distribution and the time series of incidence data up to time $t$
\begin{align}
&p_{R|I_{\leq t}, W}
(r|i_{\leq t}, w) \nn\\
&\propto p_R(r)\cdot p_W(w)\, 
\prod_{k=t-\tau}^t 
\left[\sum_{u_k\geq 0} p_{T_k,U_k|I_{<k},W,R}
(i_k + u_k, u_k | i_{<k}, w,r)\right]\nn\\
&= p_R(r)\cdot p_W(w)\, 
\prod_{k=t-\tau}^t 
\left[\sum_{u_k\geq 0} p_{T_k| U_k, I_{<k}, W,R}(i_k + u_k | u_k, i_{<k}, w,r)\right.\nn\\
&\hspace{5cm} \cdot 
p_{U_k | I_k, I_{<k}, W, R}(u_k|i_k, i_{<k}, w, r)
\Bigg]\nn\\
&= p_R(r)\cdot p_W(w) \,
\prod_{k=t-\tau}^t 
\left[\sum_{u_k\geq 0} p_{T_k|I_{<k}, W,R}(i_k + u_k | i_{<k}, w,r)\cdot 
p_{U_k | I_k}(u_k|i_k)
\right].
\label{eq:postR1}
\end{align}
The sum over the undetected cases $u_k$ is ideally running up to the total population minus the observed cases $i_k$ (neglecting effects of acquired immunity), but 
in practice it is cutoff much earlier by the distribution $p_{U_k|I_k}$,
as discussed below.

The only assumptions made to derive 
the posterior probability density  in Eq. (\ref{eq:postR1})  have been that $W$ is independent of $R$ and that $U_k$ only depends on $I_k$.
So, Eq.~(\ref{eq:postR1})  quite generally describes how to incorporate the effect of undetected cases in a Bayesian statistical model for the effective reproduction number, given the serial interval and incidence data. 

We now turn to formulate our assumptions about 
each of the terms appearing in Eq.~(\ref{eq:postR1}),
and we will reach a simple analytical form, ready to use for numerical simulations.

The prior $p_R(r)$ for $R$ is assumed to be uniform.

For the prior distribution $p_W(w)$ of the serial interval variable $W$ we assume a continuous Gamma distribution (to be evaluated on integer values of $w$)
described by two parameters: the shape parameter $a$ and the rate parameter $b$
\be
p_{W|a,b}(w|\alpha,\beta)=\frac{\beta^\alpha}{\Gamma(\alpha)}w^{\alpha-1}e^{-\beta w}\,.
\label{eq:gammadistrib}
\ee
The parameter values at $1\sigma$ level have been reported in Ref.~\cite{lombardy} as
\be
a = 1.87\pm 0.26\,,\qquad 
b = 0.28\pm 0.04\,.
\label{eq:gammadistrib1}
\ee
The prior distributions of the serial interval parameters $a, b$ are considered normal: $a\sim \mathcal{N}(1.87, 0.26^2)$, 
$b\sim \mathcal{N}(0.28, 0.04^2)$.

The probability mass function of the undetected cases $U_t$, 
 which we already assumed to depend on  $I_t$ only, 
 can be modelled in many different ways, for example with decreasing probabilities associated to large values of undetected cases, and also in a time-dependent way. 
In this paper we adopt the simplest assumption of a discrete uniform distribution with a single parameter $C_U$: 
$U_t\sim$ Uniform$([0,C_U I_t])$.
This way, we are describing the situation where the number of undetected cases at a given time $t$ can be at most $C_U$ times the number of reported cases at the same time.
We leave the investigation of alternative (and more realistic) scenarios to future work. 
So, the probability mass function of the number of undetected cases, conditioned on the values of the incidence data and the parameter $C_U$ is
\be  
p_{U_t|I_t, C_U}(u_t|i_t, c_U)\propto\chi_{[0,c_U i_t]}(u_t)\,,
\label{eq:priorUI}
\ee 
where $\chi$ is the indicator function.
The  prior distribution of the continuous parameter $C_U$ is  assumed to be an uninformative uniform prior  between 0 and 2: $C_U\sim$  Uniform$([0,2])$.
The choice of the number 2 if of course subjective, although we believe it is a reasonable and conservative prior.

The number of new total cases $T_k$ at time $k$ within the time window $[t-\tau,t]$, given the previous incidence data, the serial interval data and the value of $R$, is assumed to be Poisson-distributed with parameter $R\Lambda_k$, 
where the total infection potential at a generic time $t$ is defined by
\be
\Lambda_t(w,i_{<t})\equiv \sum_{s=1}^t i_{t-s}w_s\,.
\label{eq:Lambda}
\ee
At first approximation, 
this quantity can be considered as unaffected by the undetected cases, as the serial interval distribution is derived from tracking the secondary of reported cases. By considering the sample of secondary infections as representative of the population, the approximation above is justified. The generalizations to replace 
the Poisson distribution with a two-parameter negative binomial distribution and
to include the undetected cases into $\Lambda_t$ are left to future work.

Therefore, we can write explicitly the 
probability mass function of ${T_k|I_{<k},W,R}$ as
\be
p_{T_k|I_{<k},W,R}(i_k+u_k|i_{<k},w,r)=
\frac{1}{(i_k+u_k)!}e^{-r\Lambda_k(w, i_{<k})}
\left[r\Lambda_k(w, i_{<k})\right]^{i_k+u_k}\,.
\ee 
Now we have collected all the ingredients of 
Eq.~(\ref{eq:postR1}), and we added the nuisance parameters $a,b,C_U$ to the model. So the joint posterior density of $R$ and the nuisance parameters reads
\begin{align}
p_{R,a,b,C_U|I_{\leq t}, W}
(r,\alpha,\beta,c_U|i_{\leq t}, w)
&\propto p_R(r)\, p_{W|a,b}(w|\alpha,\beta)
\nn\\
&\hspace{-3cm}\times 
\prod_{k=t-\tau}^t
\left[
\sum_{u_k\geq 0}
e^{-r\Lambda_k(w, i_{<k})}\frac{
\left[r\Lambda_k(w, i_{<k})\right]^{i_k+u_k}}{(i_k+u_k)!}
\cdot p_{U_k | I_k, C_U}(u_k|i_k, c_U)
\right]\,.
\label{eq:postR2}
\end{align} 
By marginalizing over $a,b,C_U$ we finally get the posterior probability density for $R$
\begin{align}
p_{R|I_{\leq t}, W}
(r|i_{\leq t}, w)
&\propto 
p_R(r)\int d\alpha\,d\beta\,dc_U\,p_a(\alpha)\, p_b(\beta)\, p_{C_U}(c_U)\, p_{W|a,b}(w|\alpha,\beta)\nn\\
&\hspace{-1cm} \times 
\prod_{k=t-\tau}^t
\left[
\sum_{u_k\geq 0}
e^{-r\Lambda_k(w, i_{<k})}\frac{
\left[r\Lambda_k(w, i_{<k})\right]^{i_k+u_k}}{(i_k+u_k)!}
\cdot p_{U_k | I_k, C_U}(u_k|i_k, c_U)
\right]\,.
\label{eq:postR3}
\end{align} 
With a uniform prior for the number of undetected cases
as described in Eq.~(\ref{eq:priorUI}),
the sum in the square bracket of Eq.~(\ref{eq:postR3})
can be computed analytically in terms of
the regularized upper incomplete gamma function $Q(s,x)$ 
as
\begin{align}
p_{R|I_{\leq t}, W}
(r|i_{\leq t}, w)
&\propto p_R(r)\, 
\int d\alpha\, d\beta\,dc_U\, p(\alpha)\, p(\beta)
\,p_{C_U}(c_U)\, 
p_{W|a,b}(w|\alpha,\beta) \nn\\ 
& \times
\prod_{k=t-\tau}^t
\left[
 Q\left((1+c_U)i_k+1,r\Lambda_k(w, i_{<k})\right)
 -Q\left(i_k,r\Lambda_k(w, i_{<k})\right)
 \right]
 \,.
 \label{eq:postR4}
\end{align} 
At any time $t>\tau$, 
from this conditional posterior probability density it is possible
to compute the mean and the  95\% central credible intervals. In particular, the effective reproduction number $R_t$ is the
expected value of $R$ conditioned on the past incidence data and serial interval
\be 
R_t=\textrm{E}[R|I_{\leq t},W]=\int dr\, r\, p_{R|I_{\leq t}, W}
(r|i_{\leq t}, w)\,,
\ee 
where the  conditional probability density $p_{R|I_{\leq t}, W}
(r|i_{\leq t}, w)$
marginalized over all nuisance parameters is given by Eq.~(\ref{eq:postR3}) in general, and by Eq.~(\ref{eq:postR4}) for the particular case of uniform distribution of the number of undetected cases.

\subsection{Incidence data by country}
\begin{figure}[t] 
    \centering
    \vspace{-0.5cm}
    \includegraphics[width=0.4\textwidth]{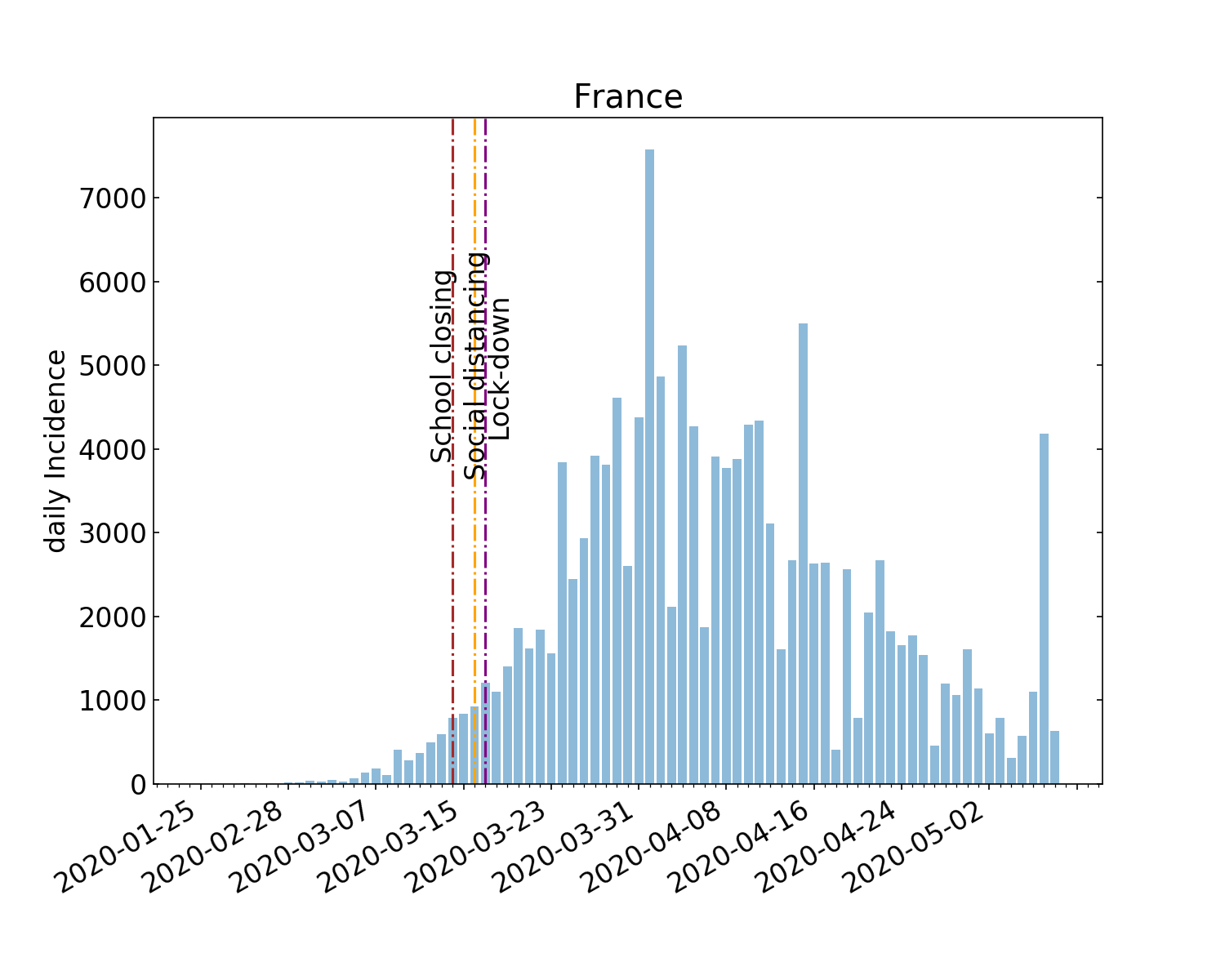}
    \includegraphics[width=0.4\textwidth]{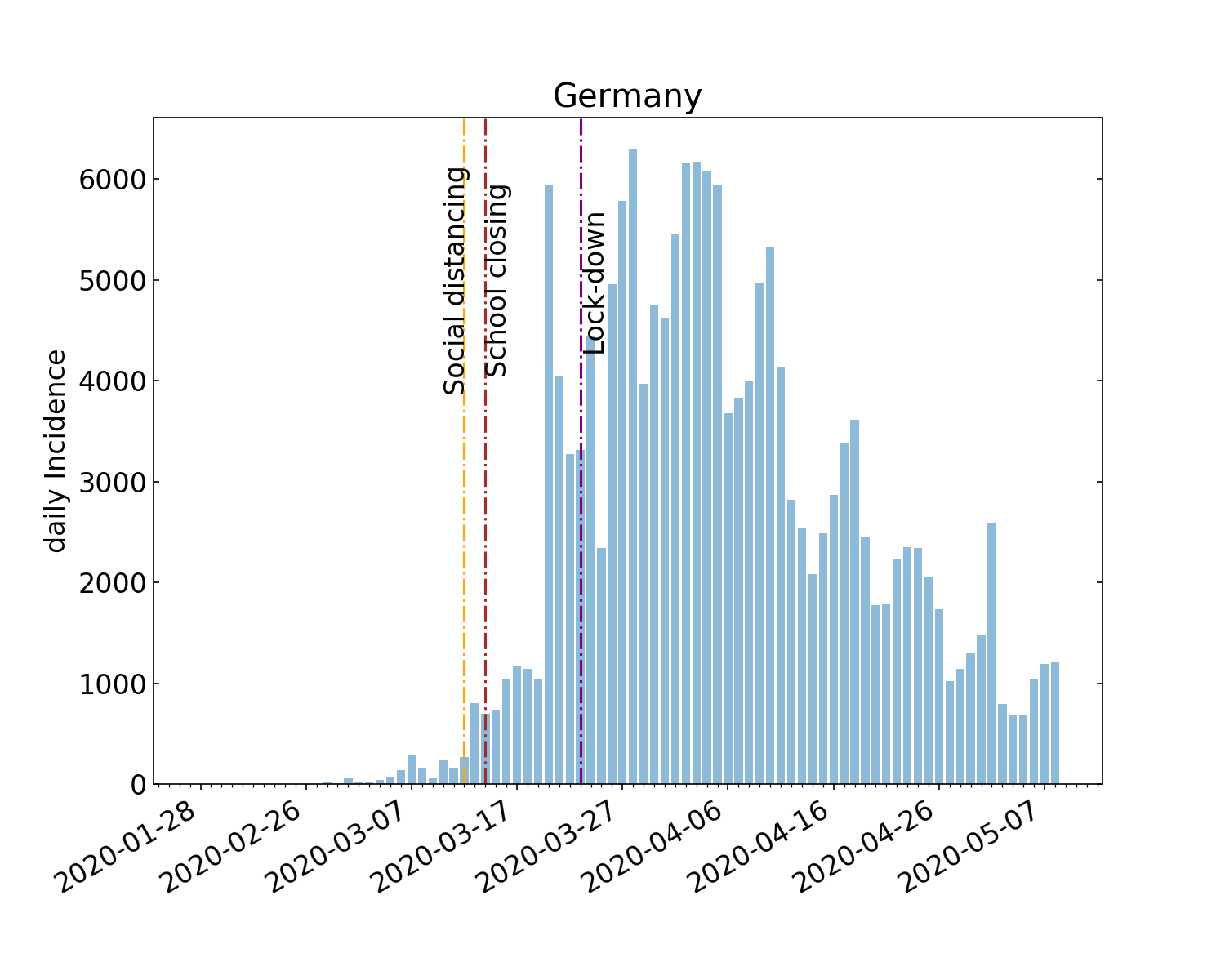}\\
    \vspace{-0.5cm}
    \includegraphics[width=0.4\textwidth]{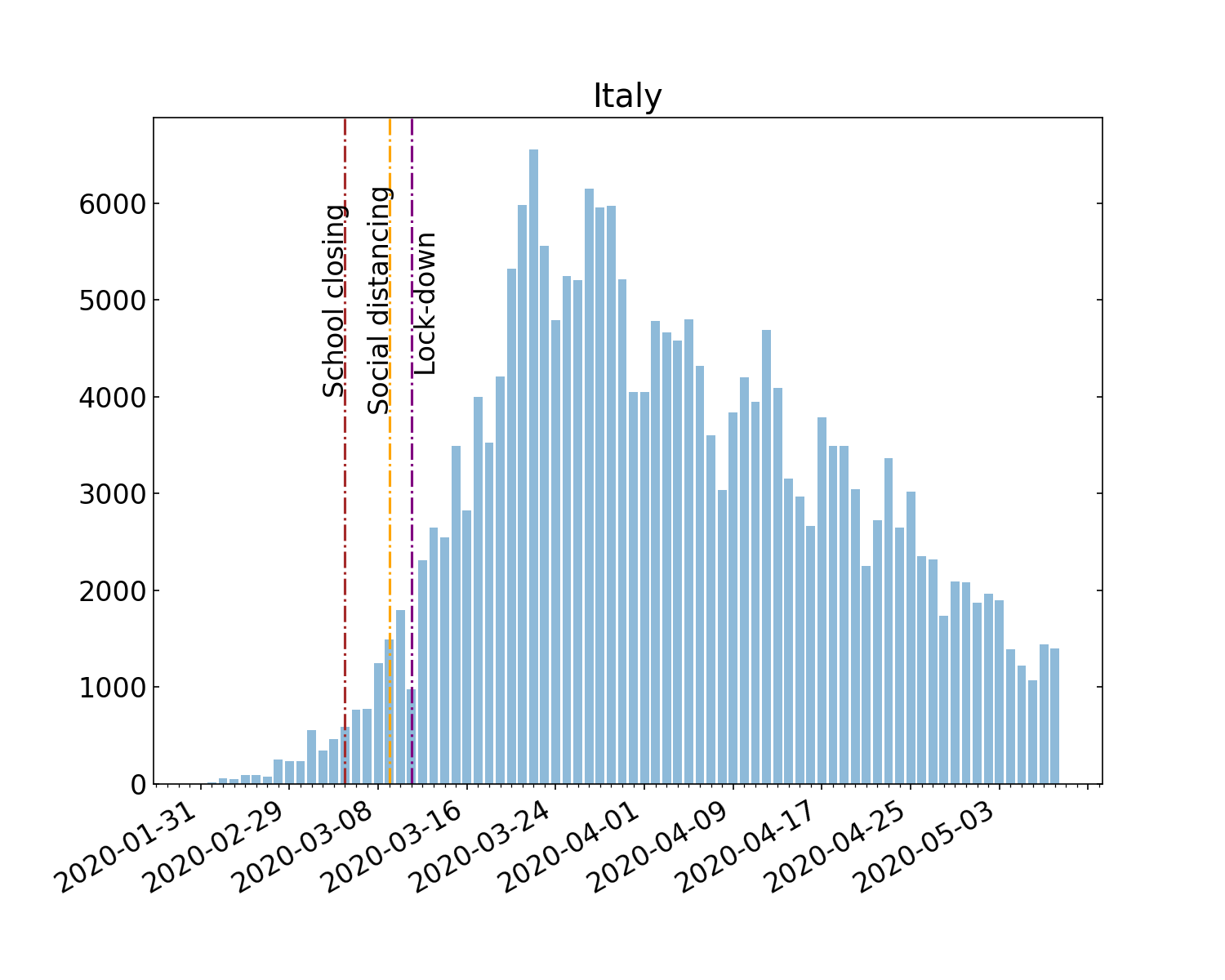}
    \includegraphics[width=0.4\textwidth]{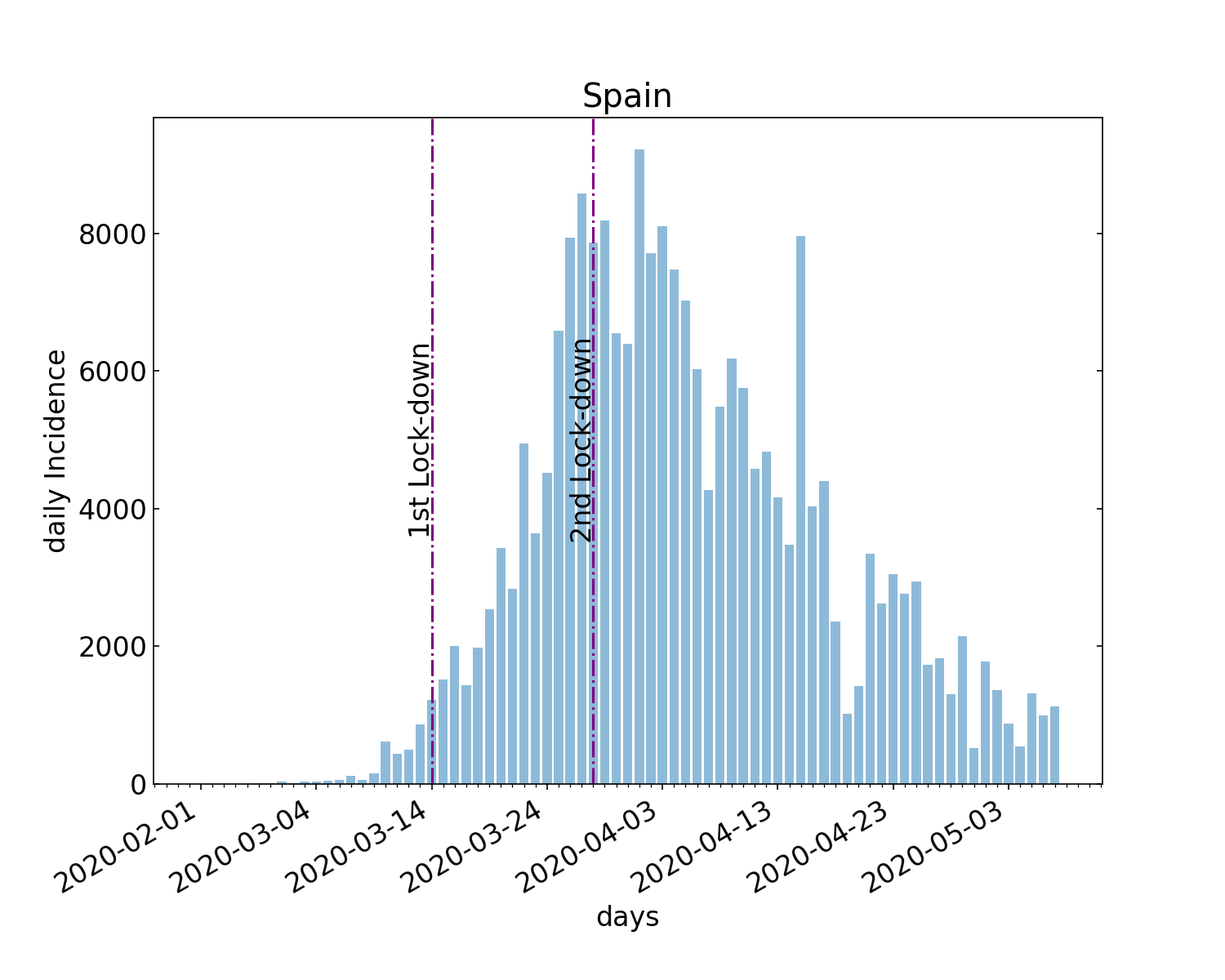}\\
    \vspace{-0.5cm}
    \includegraphics[width=0.4\textwidth]{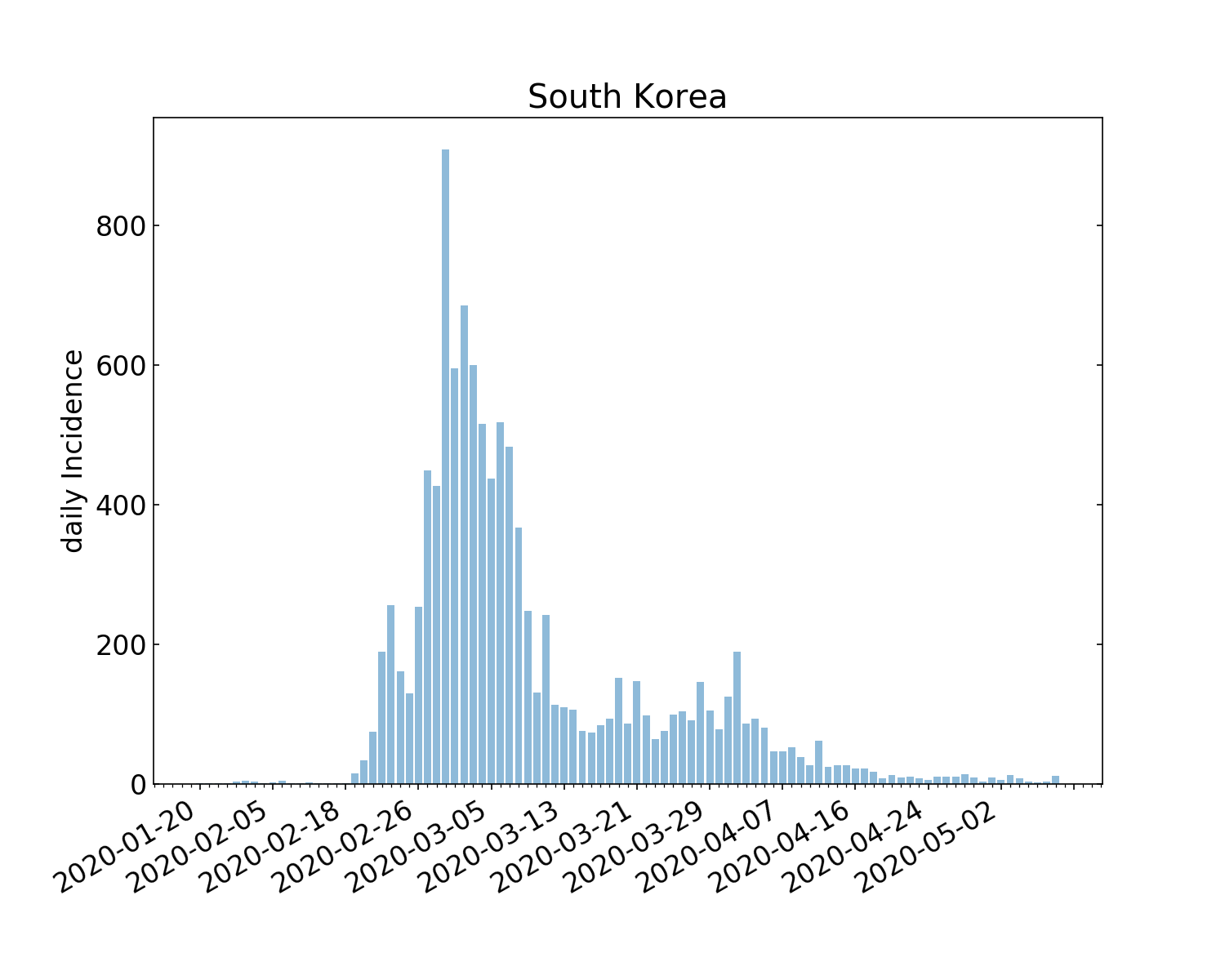}
    \includegraphics[width=0.4\textwidth]{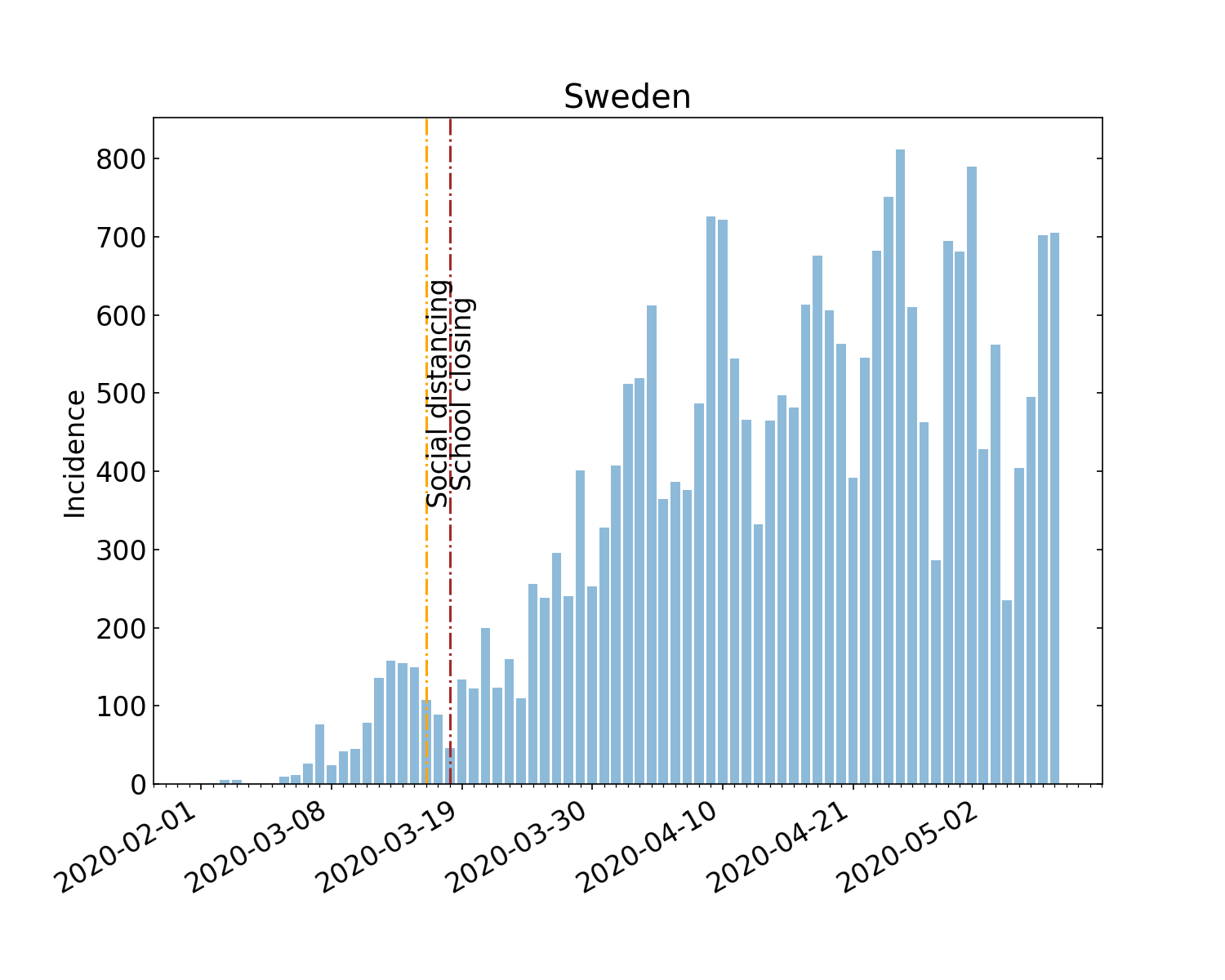}\\
    \vspace{-0.5cm}
    \includegraphics[width=0.4\textwidth]{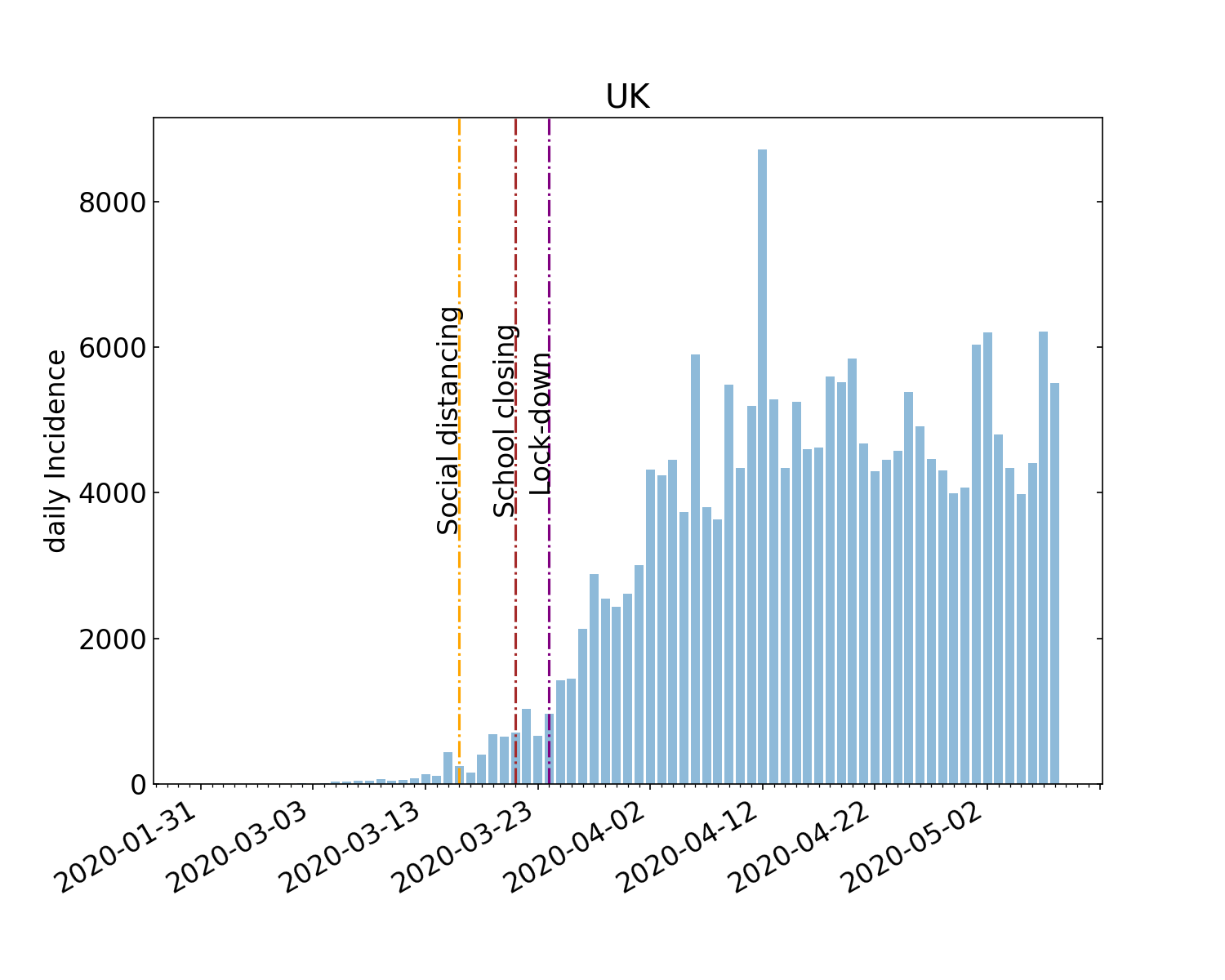}
    \includegraphics[width=0.4\textwidth]{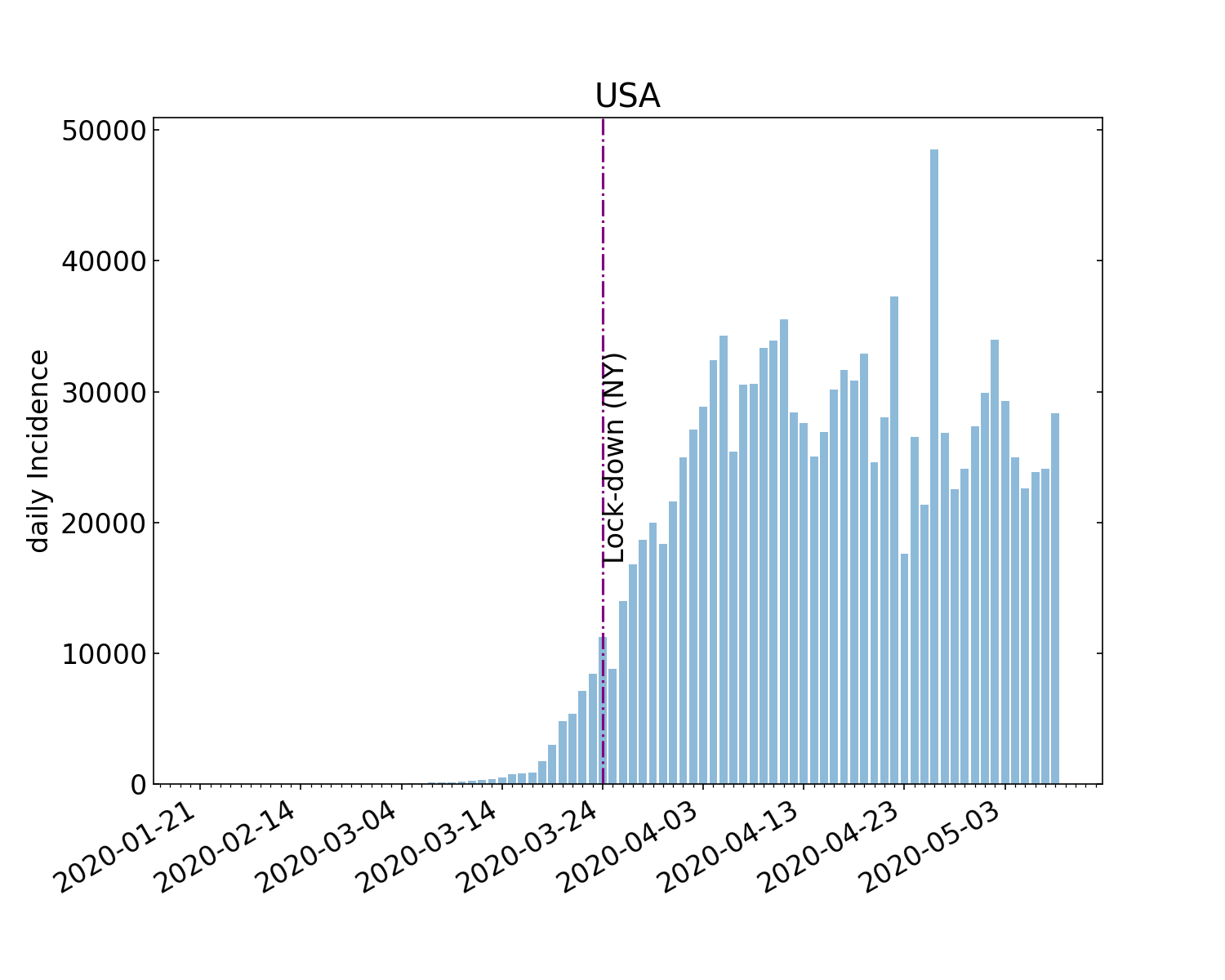}\\
\caption{\small \emph{Number of new daily infection cases of COVID-19 in the countries considered in this paper. The vertical lines refer to the time when containment measures have been adopted (for the USA, we show the lock-down date in the state of NY as a reference). 
}}
\label{fig:Incidence}
\end{figure}

The daily incidence data, i.e. the number of new positive cases on each day, we used for the plots in Section \ref{sec:methods} are plotted in figure~\ref{fig:Incidence}. Notice how the pattern of disease incidence in Italy, Spain, France, Germany, UK and USA are different from the pattern in South Korea and Sweden. 
For Spain, a negative incidence of -1400 was reported on 2020-04-19. We believe this is due to fixing numbers which were incorrectly reported previously. In order to deal with positive incidence data only, 
we adjust the negative number of cases reported on 2020-04-19 by distributing the cases of the previous and following day.
In particular, we assign one-half of the reported cases on 2020-04-18 and
one-fourth of the cases on 2020-04-20 to the daily incidence on 2020-04-19, resulting in a total of 590 cases on 2020-04-19.

\subsection{Serial interval distribution}
\begin{figure}[t] 
    \centering
    
    \includegraphics[width=0.80\textwidth]{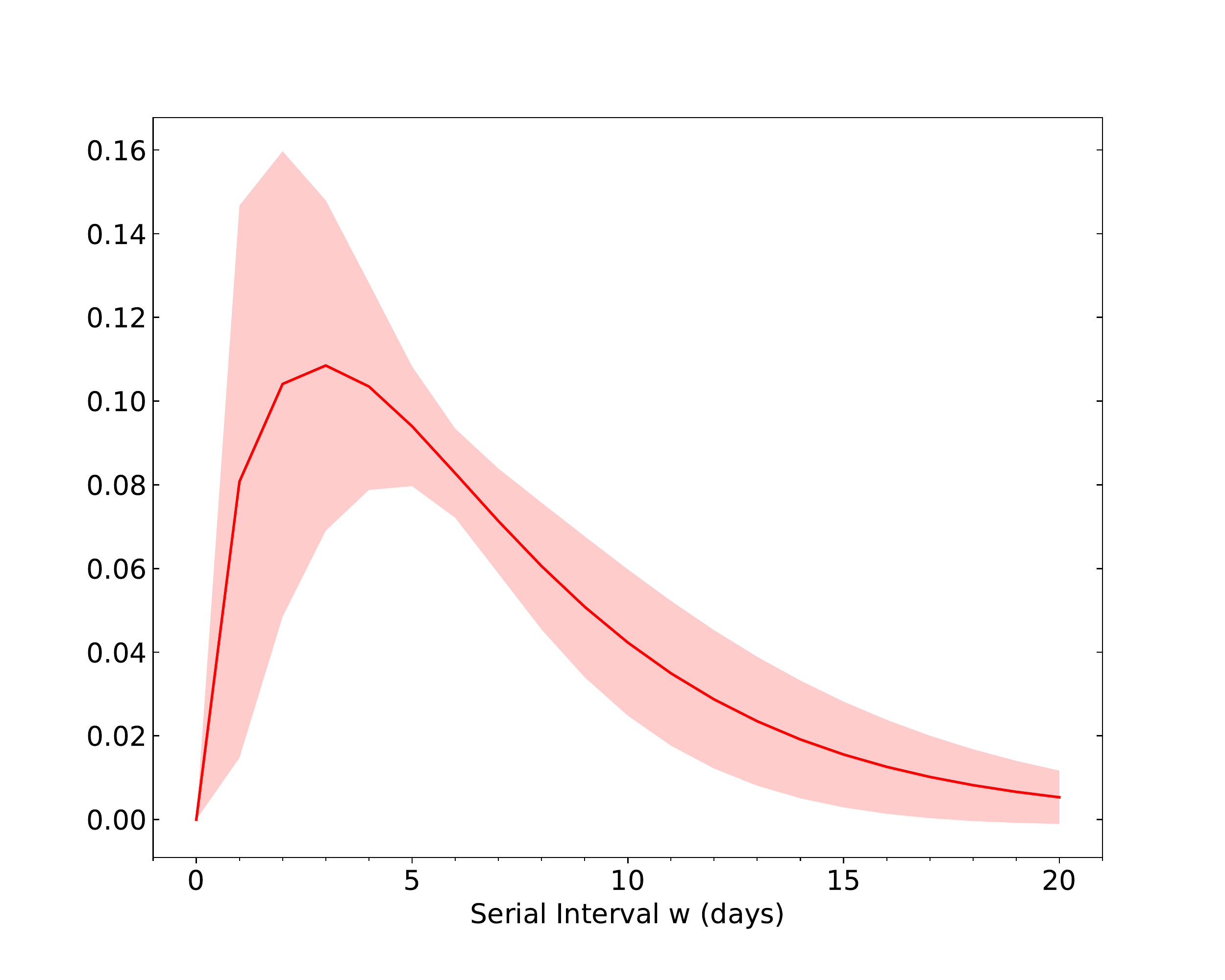}\\
\caption{{\small \emph{Serial interval distribution, 
as a gamma distribution in Eq.~(\ref{eq:gammadistrib}).
The solid line corresponds to the central values of the parameters in Eq.~(\ref{eq:gammadistrib1}). 
The shaded region is the 95\% confidence interval.}}}
\label{fig:SI}
\end{figure}

The serial interval is the number of days occurring from the onset of symptoms in a
patient and the onset of symptoms in a secondary patient  infected by the primary one.
For modelling the serial interval we used a gamma distribution (see Eq.~(\ref{eq:gammadistrib})). We simulate different realizations of the gamma distribution considering the parameters $a$ and $b$ normally distributed, with mean and variance
reported in Eq.~(\ref{eq:gammadistrib1}).
The resulting 95\% confidence interval is shown in figure~\ref{fig:SI}

\subsection{Results for the Italian regions}
\label{app:regions}
\begin{figure}[t] 
    \centering
    \includegraphics[width=0.5\textwidth]{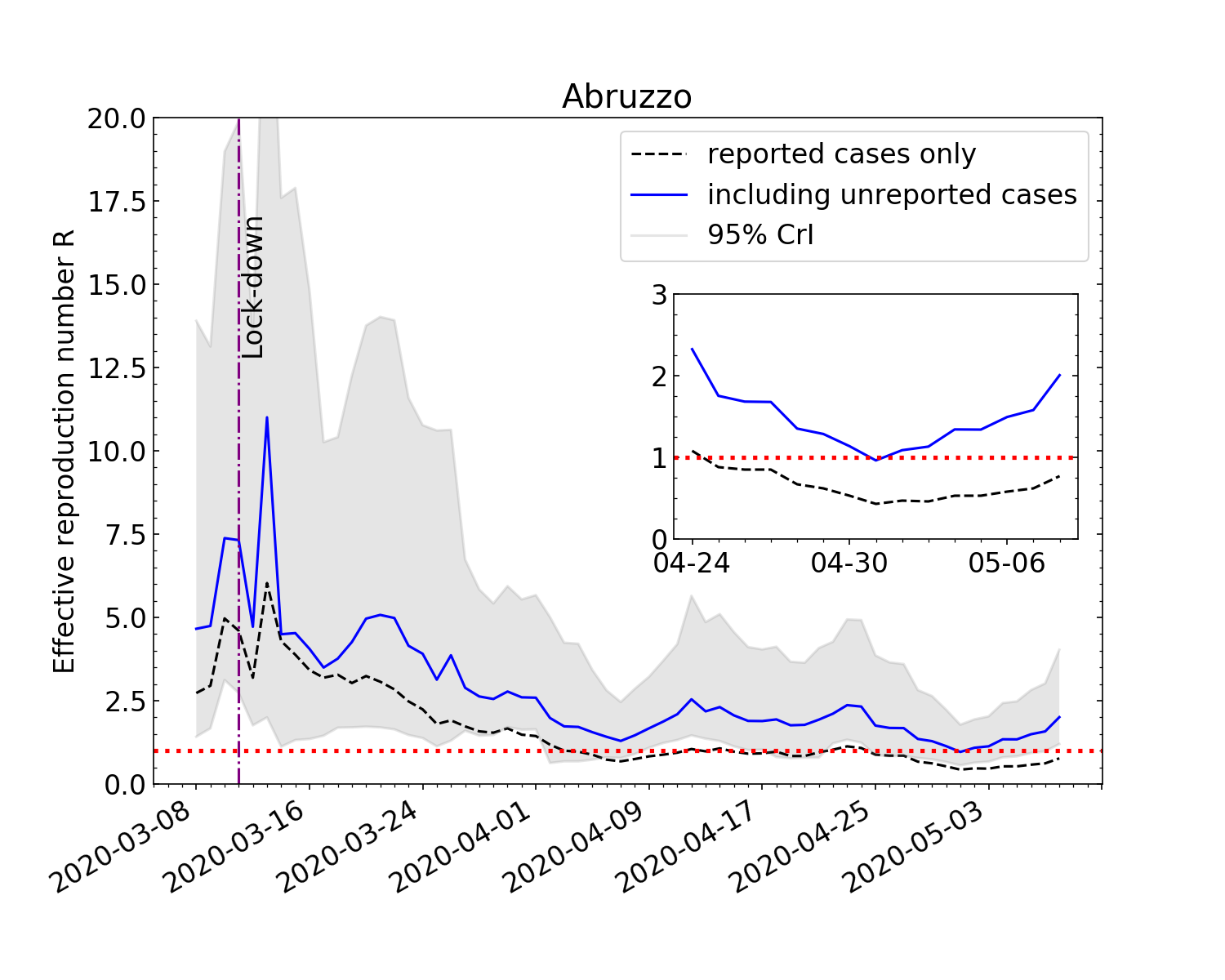}
    \hspace{-0.5cm}
    \includegraphics[width=0.5\textwidth]{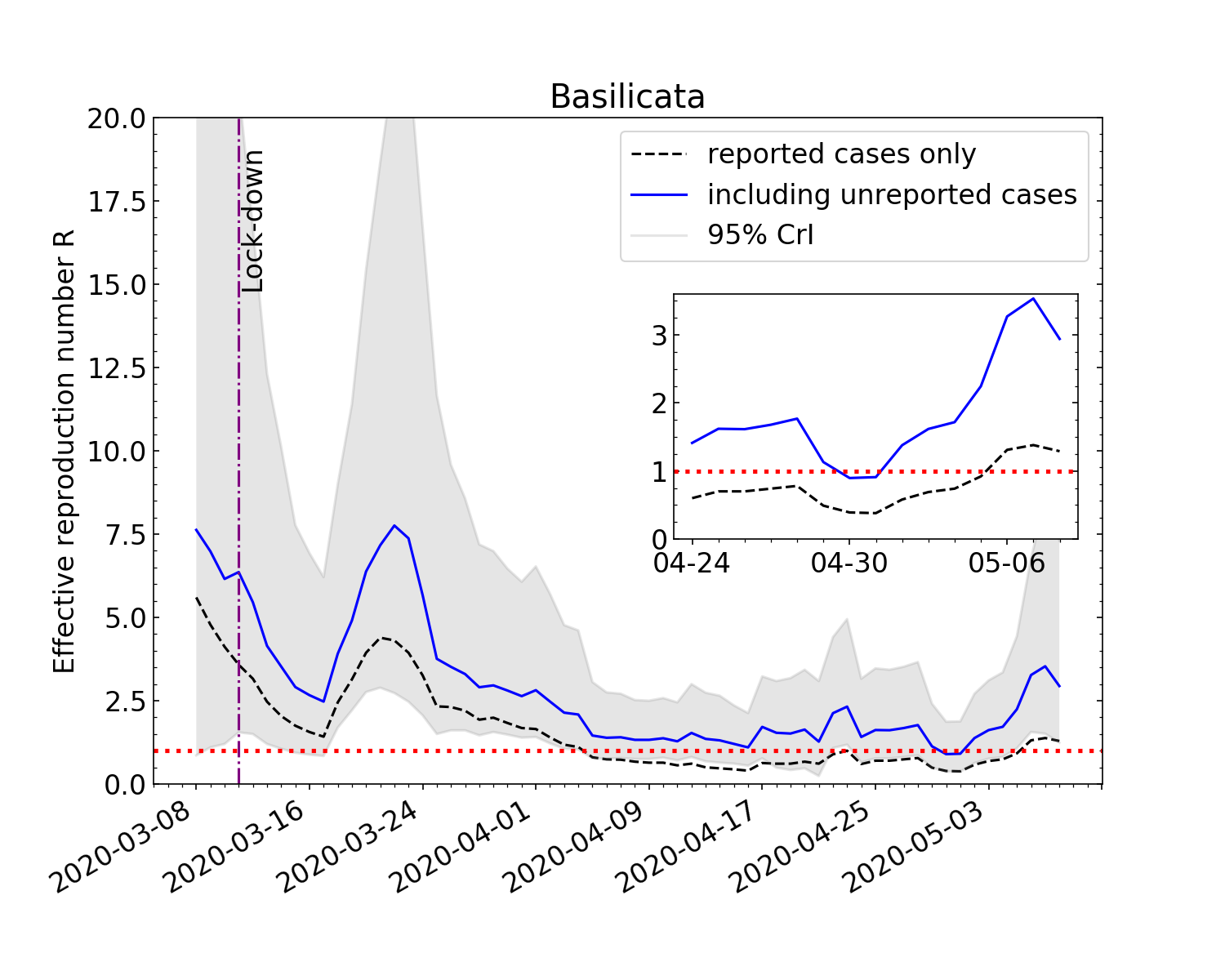}\\
    \includegraphics[width=0.5\textwidth]{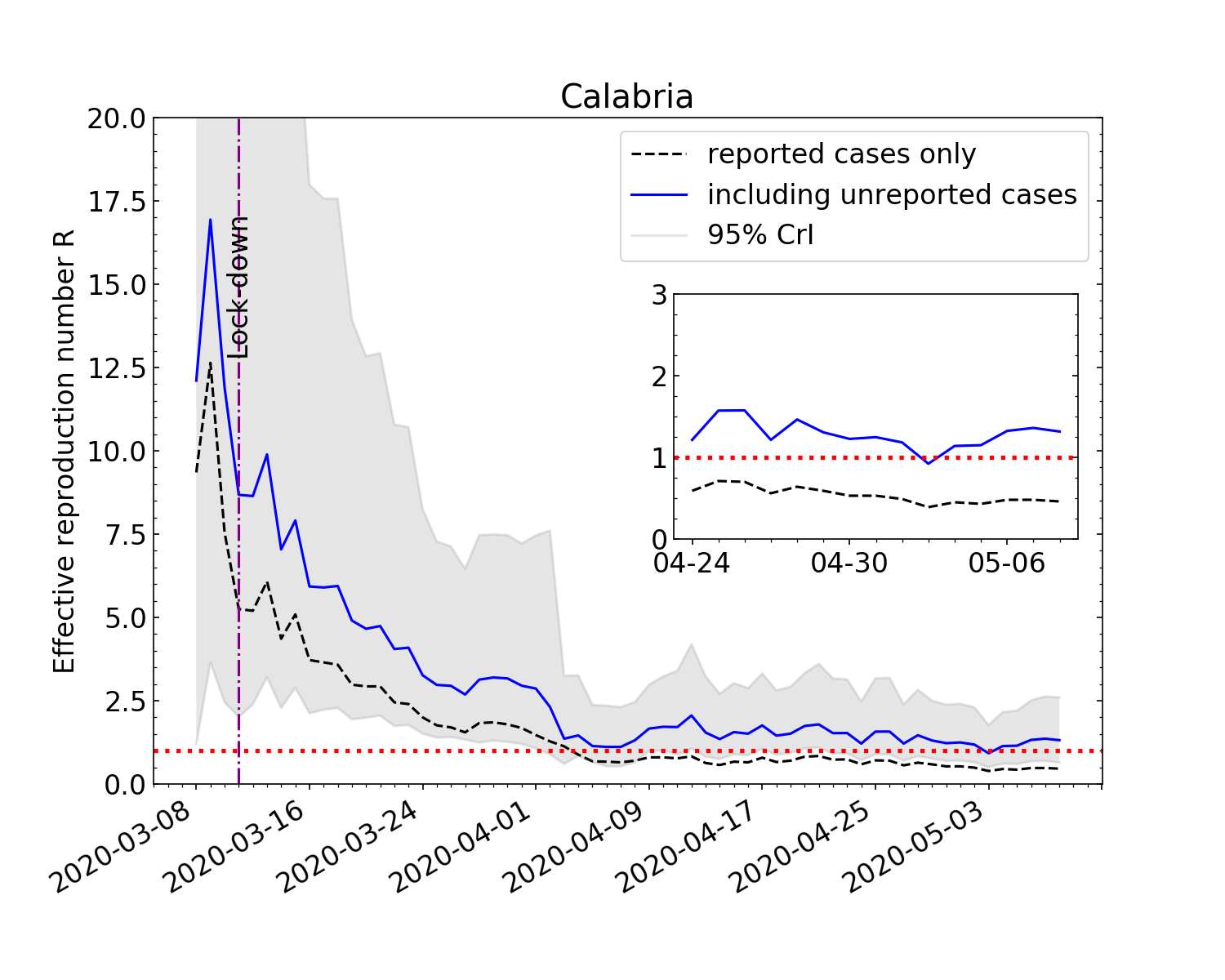}
    \hspace{-0.5cm}
    \includegraphics[width=0.5\textwidth]{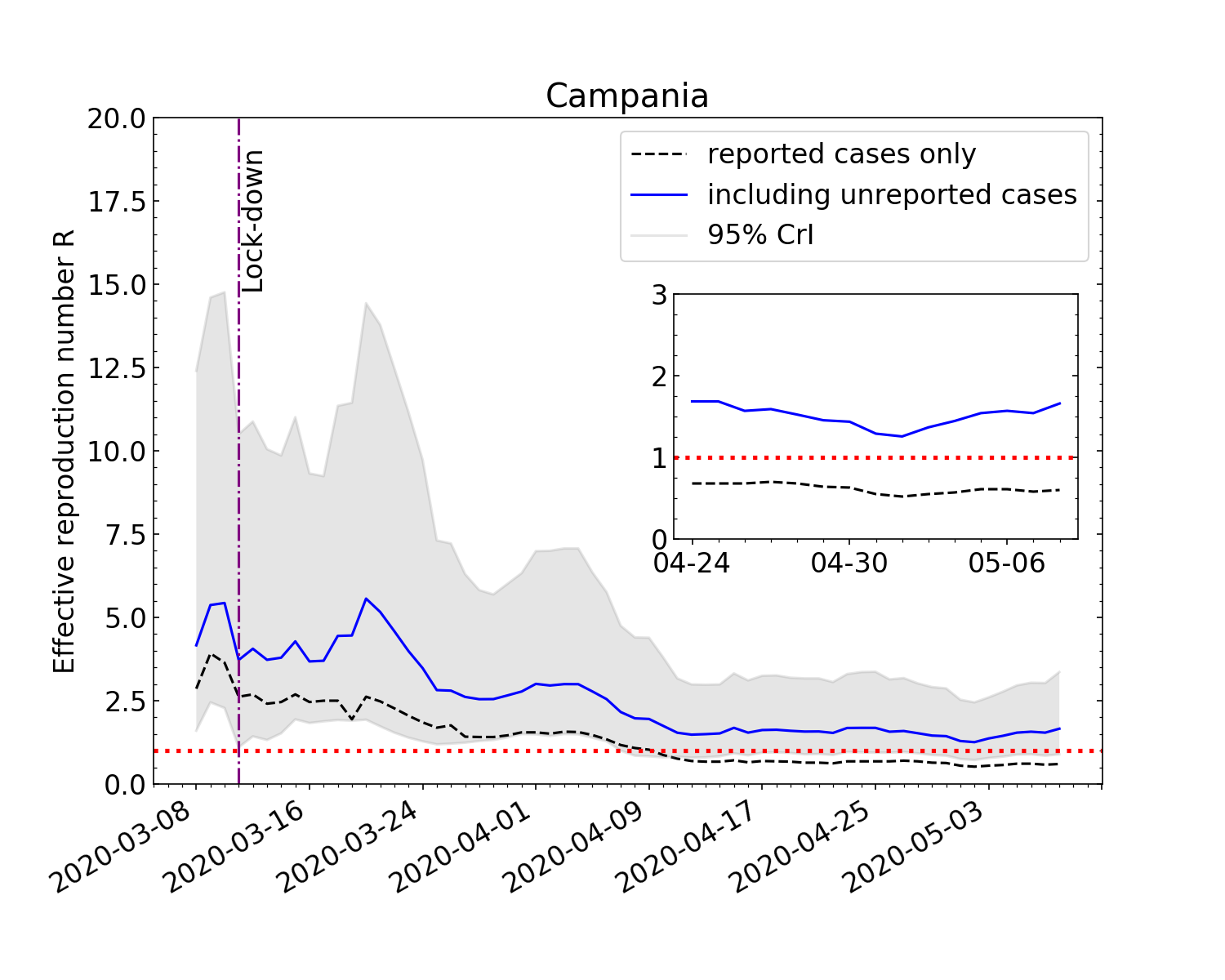}\\
    \caption{{\small 
    \emph{Evolution of effective reproduction number  for COVID-19
in four Italian regions: Abruzzo, Basilicata, Calabria, Campania.}}}
\label{fig:regions1}
\end{figure}

\begin{figure}[t] 
    \centering
    \includegraphics[width=0.5\textwidth]{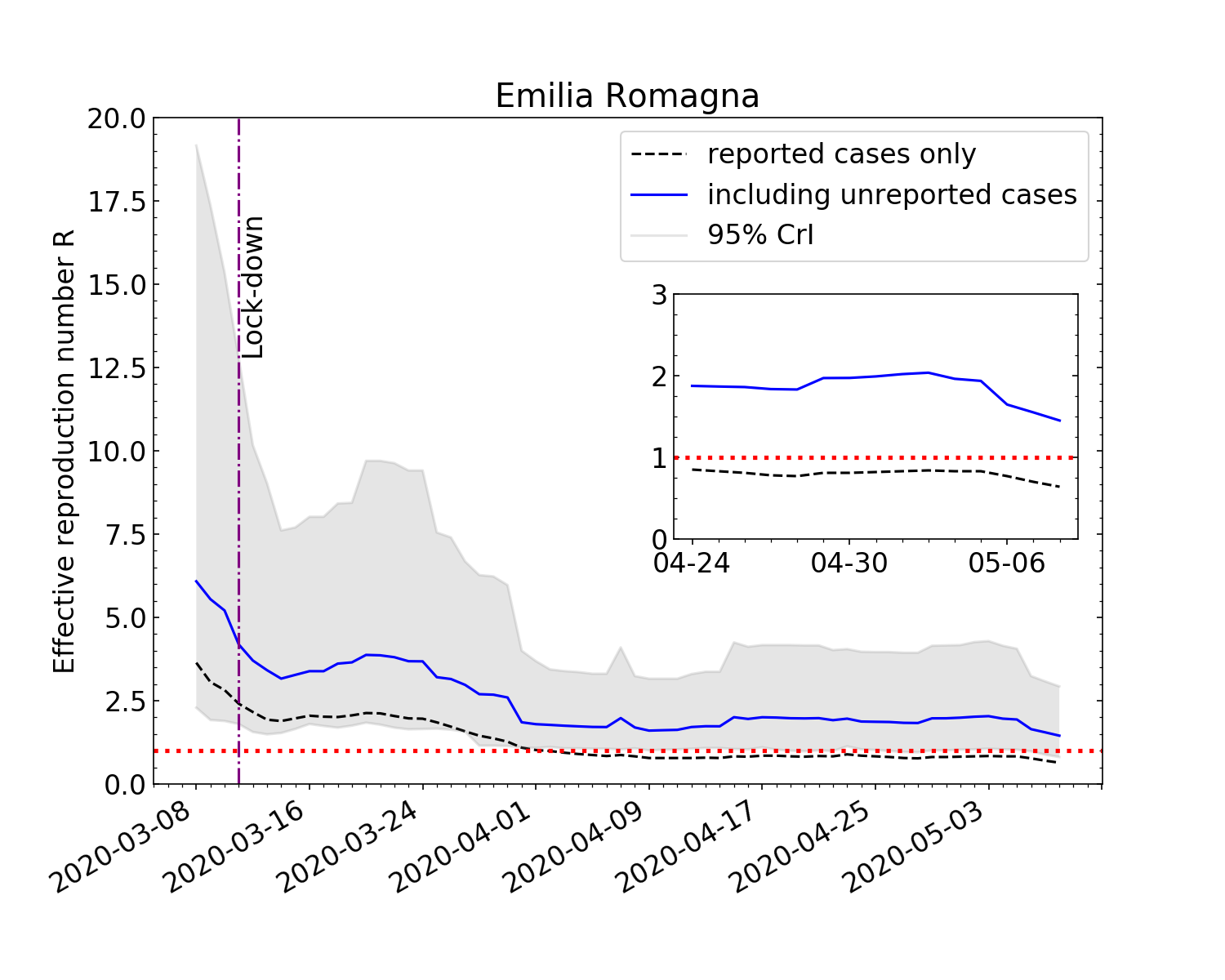}
    \hspace{-0.5cm}
    \includegraphics[width=0.5\textwidth]{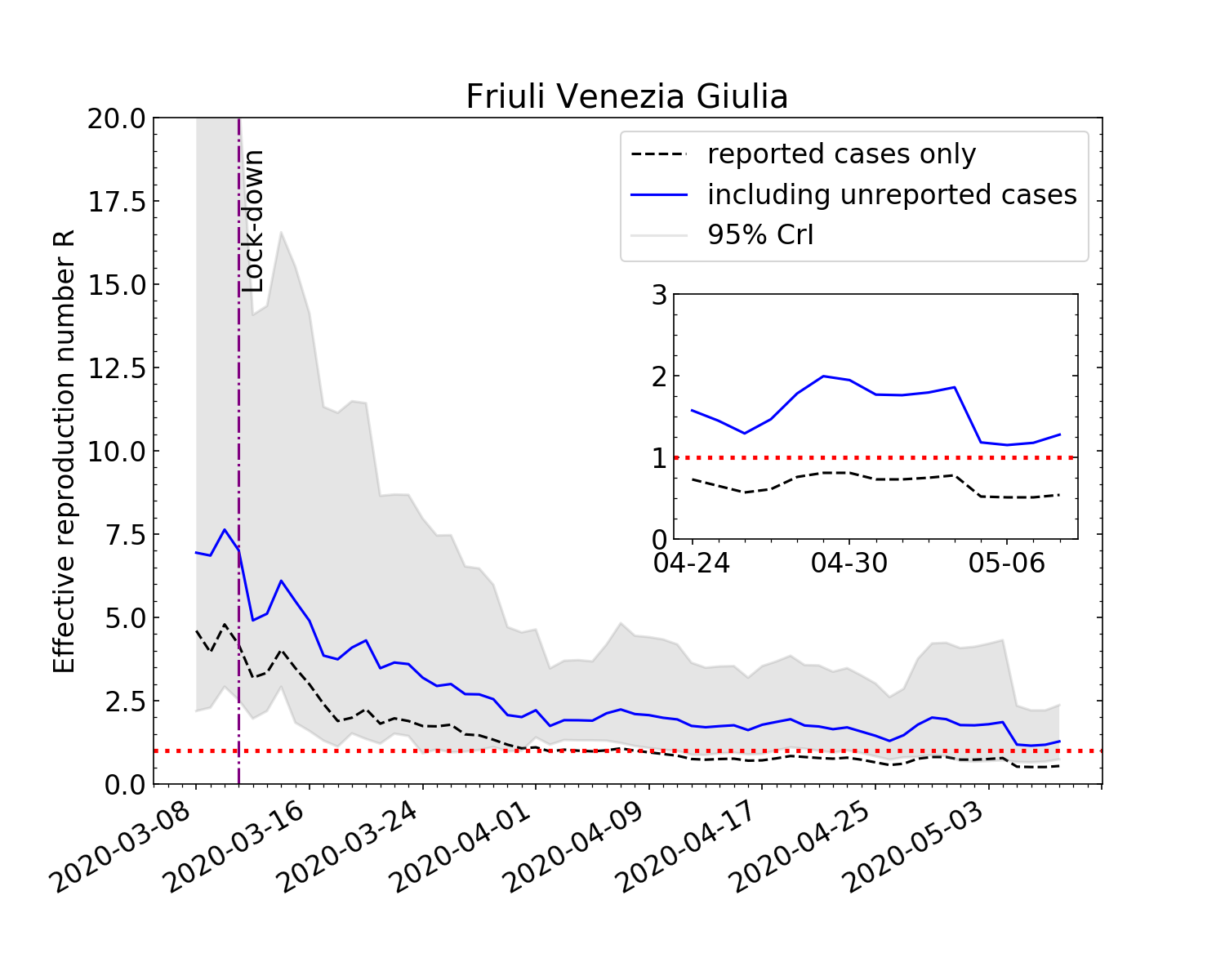}\\
    \includegraphics[width=0.5\textwidth]{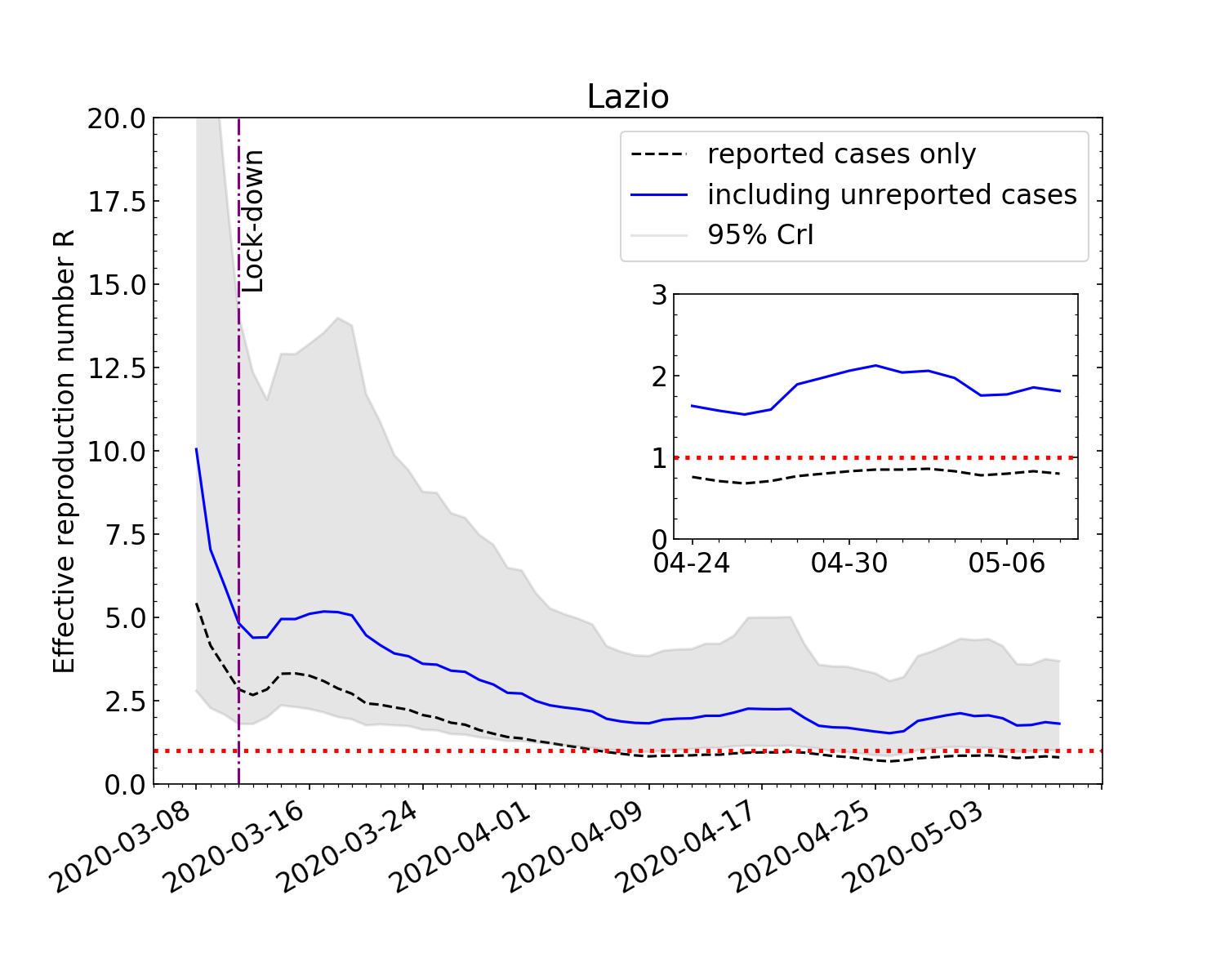}
    \hspace{-0.5cm}
    \includegraphics[width=0.5\textwidth]{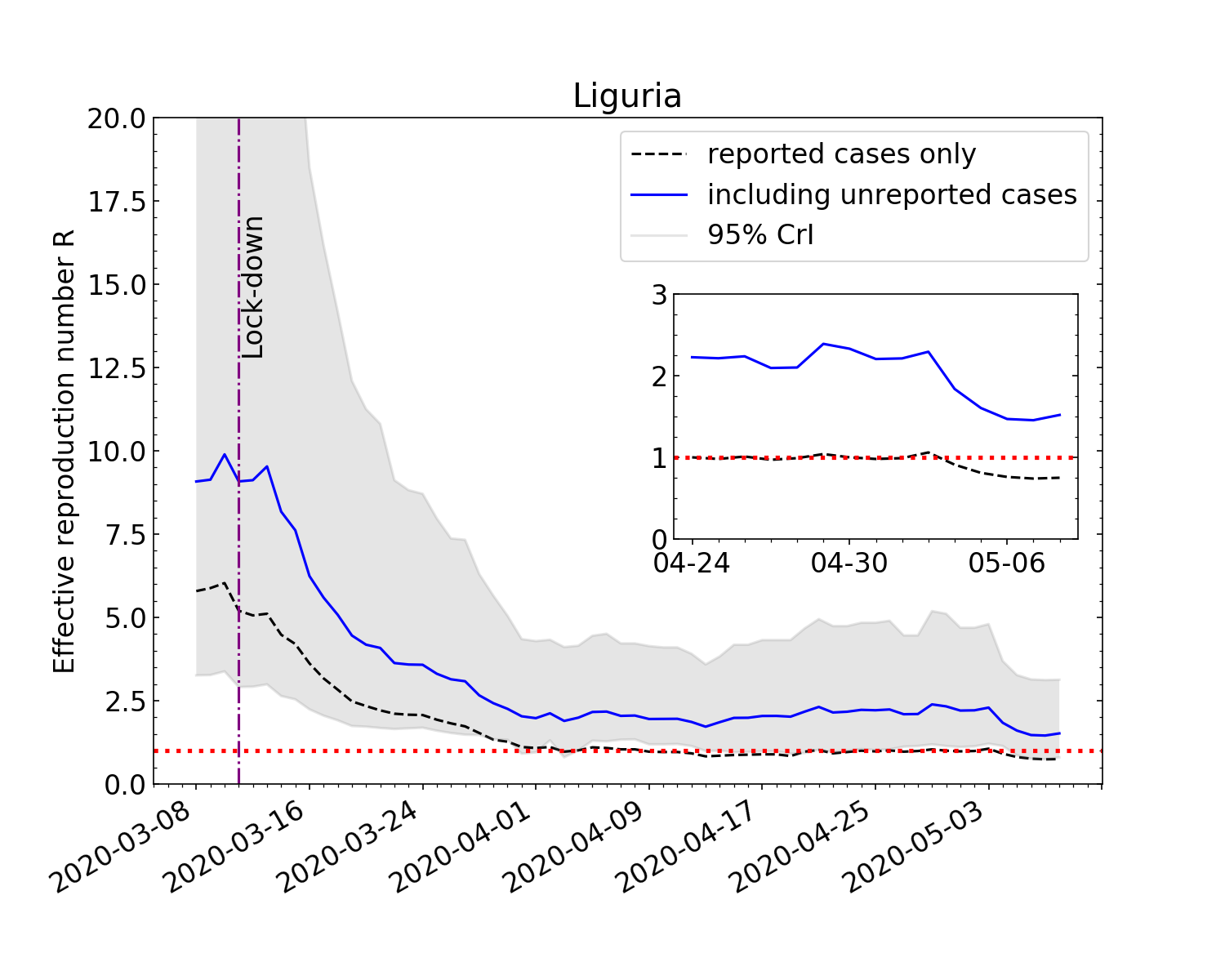}\\
\caption{{\small \emph{Evolution of effective reproduction number
for COVID-19
in four Italian regions: Emilia Romagna, Friuli Venezia Giulia, Lazio, Liguria.}}}
\label{fig:regions2}
\end{figure}

\begin{figure}[t] 
    \centering
    \includegraphics[width=0.5\textwidth]{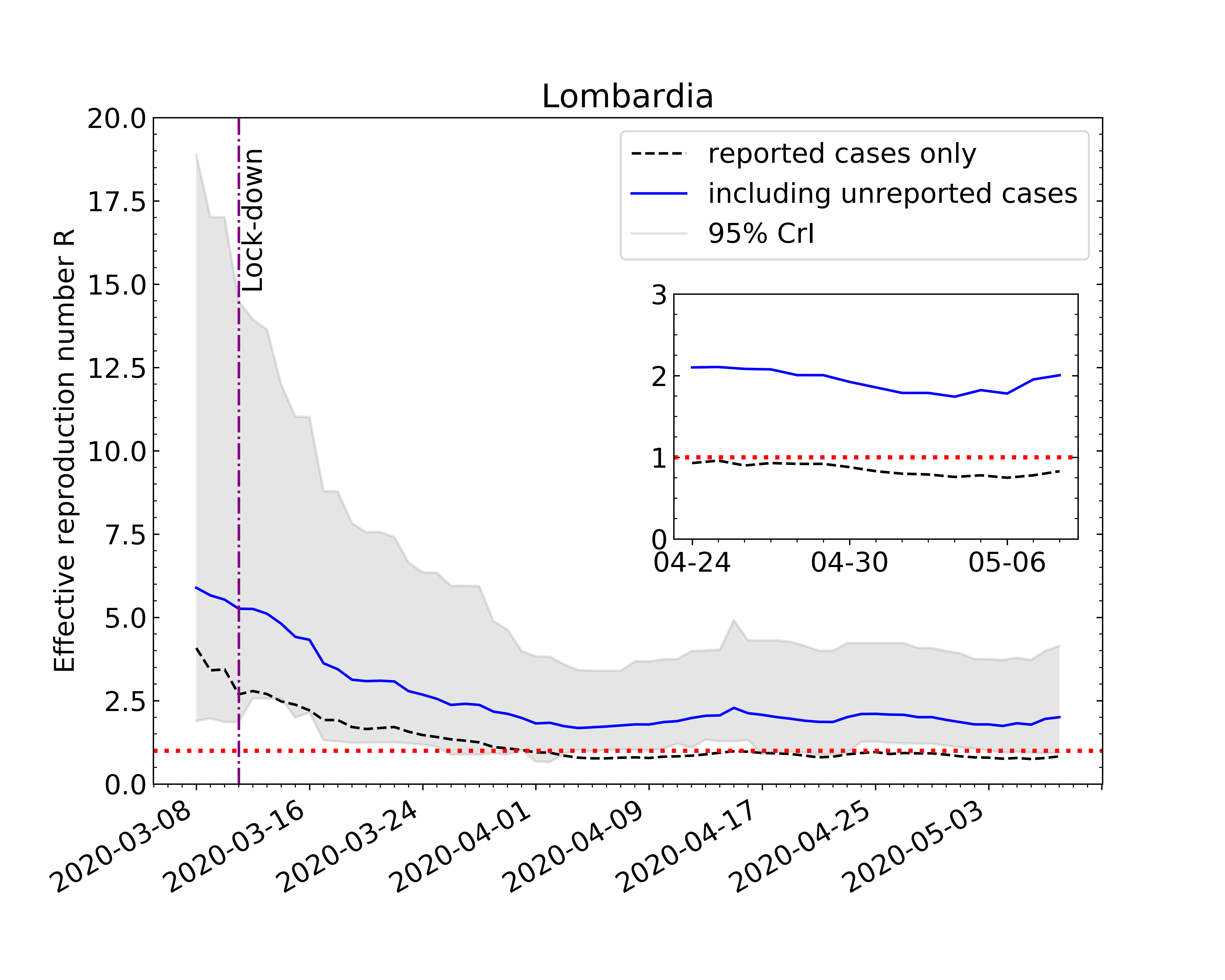}
    \hspace{-0.5cm}
    \includegraphics[width=0.5\textwidth]{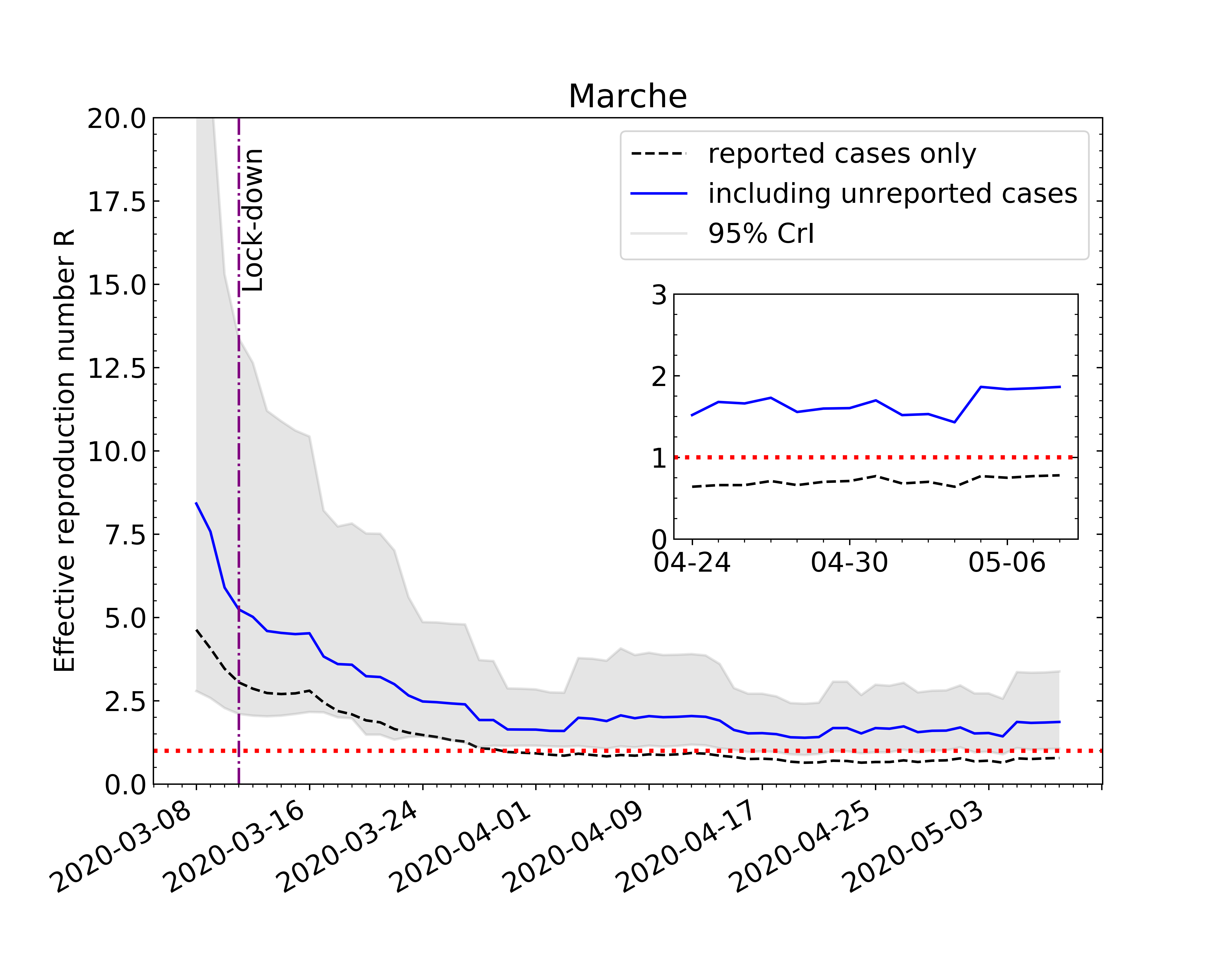}\\
    \includegraphics[width=0.5\textwidth]{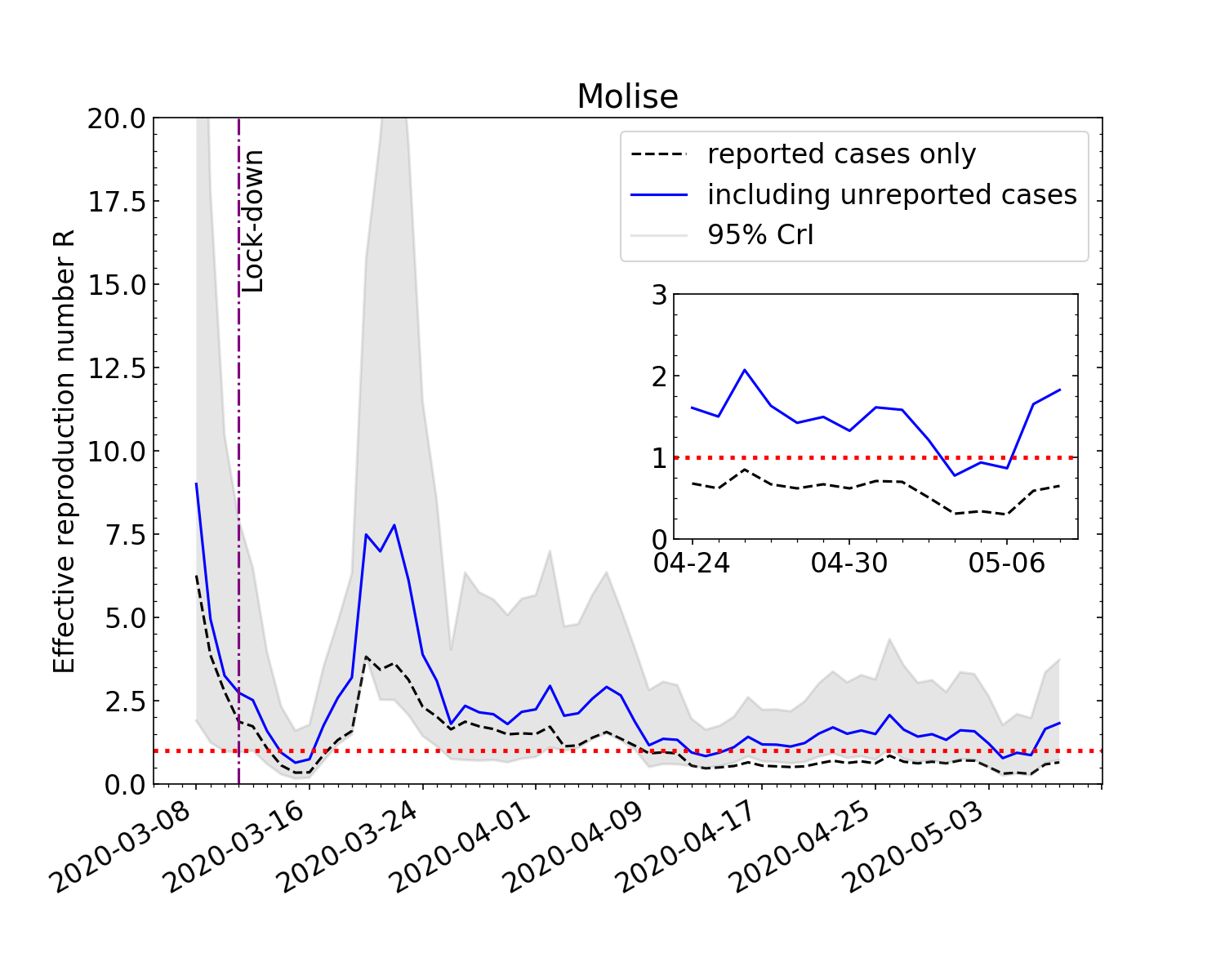}
    \hspace{-0.5cm}
    \includegraphics[width=0.5\textwidth]{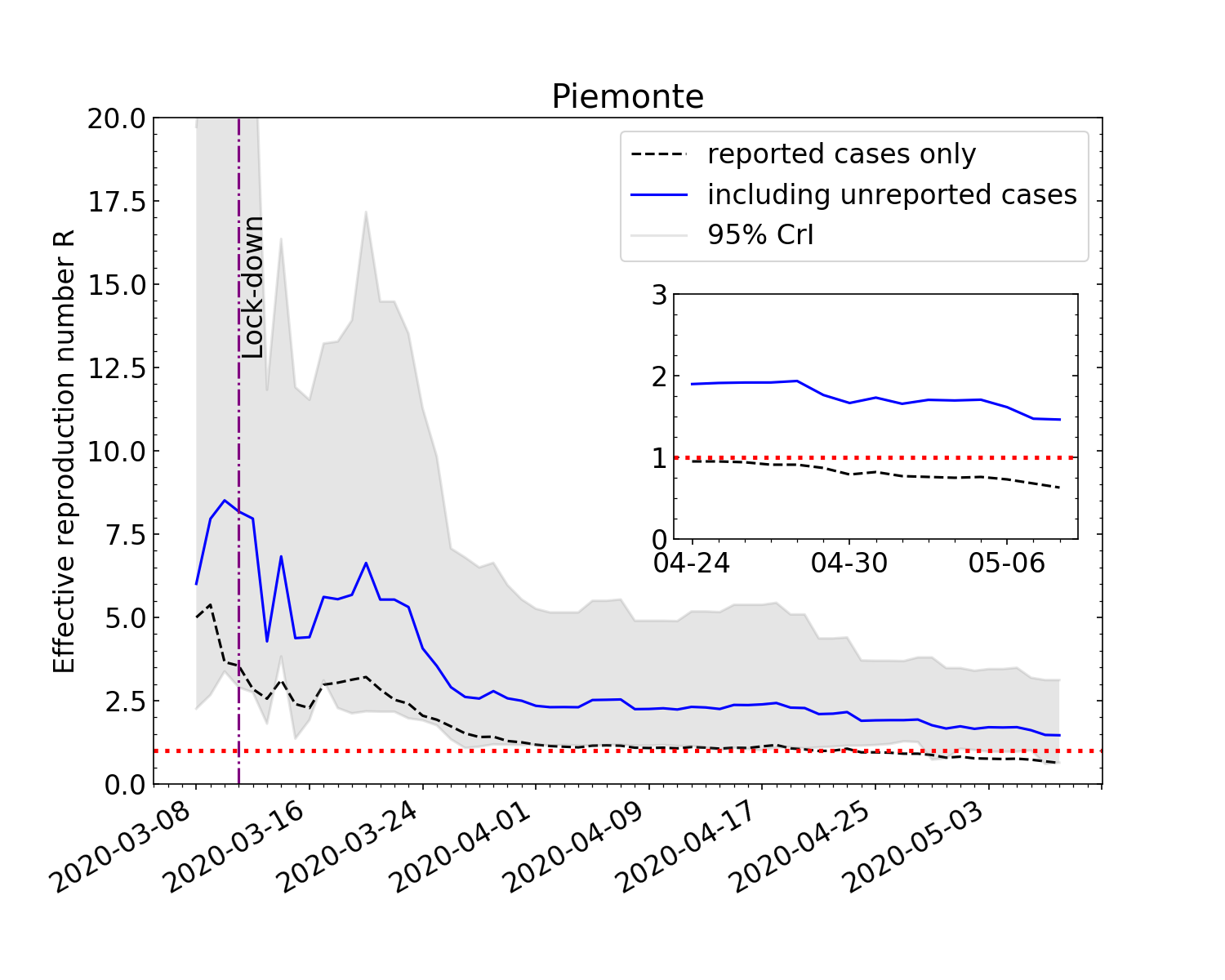}\\
    \caption{{\small \emph{Evolution of effective reproduction number 
     for COVID-19
in four Italian regions: Lombardia, Marche, Molise, Piemonte.}}}
\label{fig:regions3}
\end{figure}

\begin{figure}[t] 
    \centering
    \includegraphics[width=0.5\textwidth]{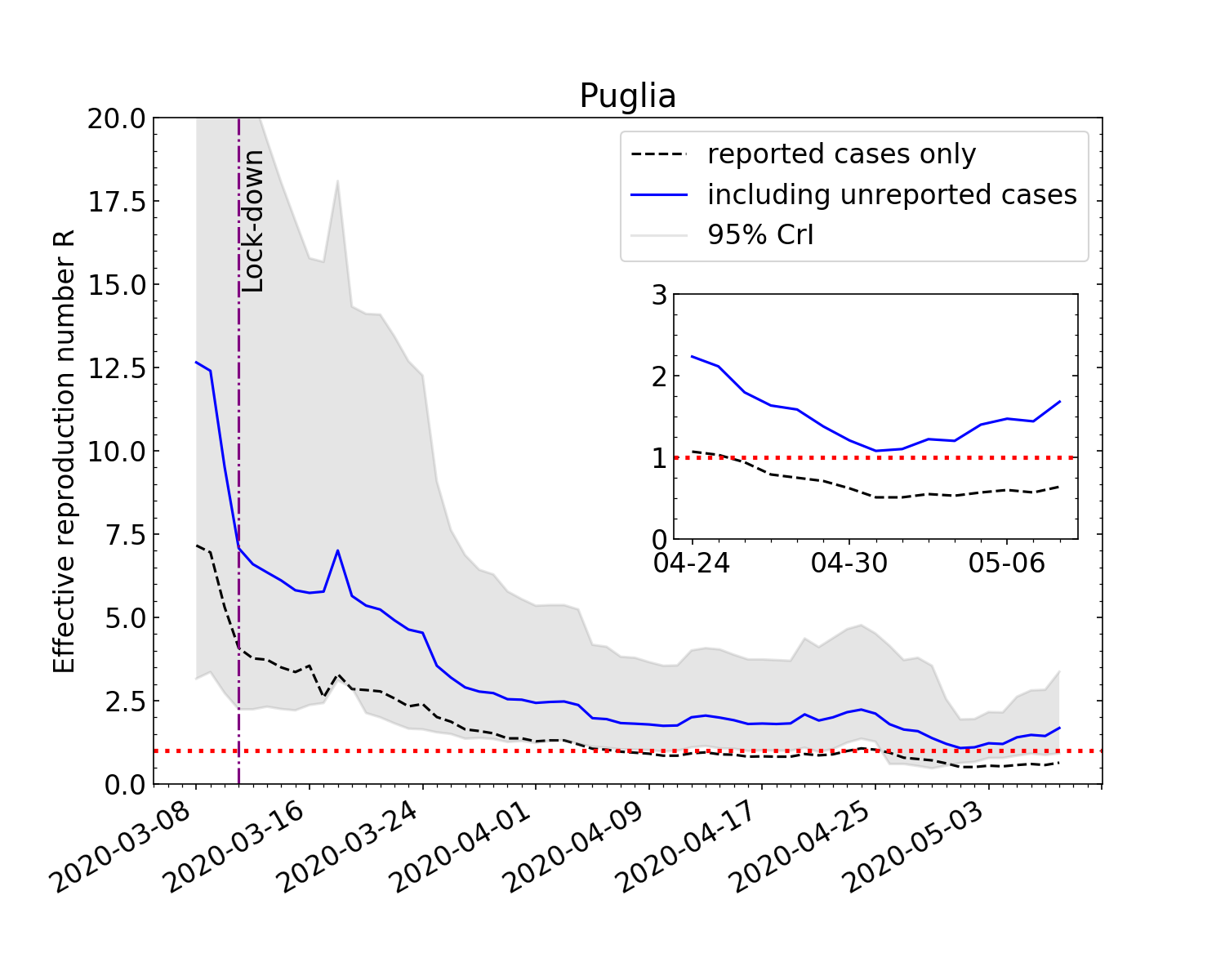}
    \hspace{-0.5cm}
    \includegraphics[width=0.5\textwidth]{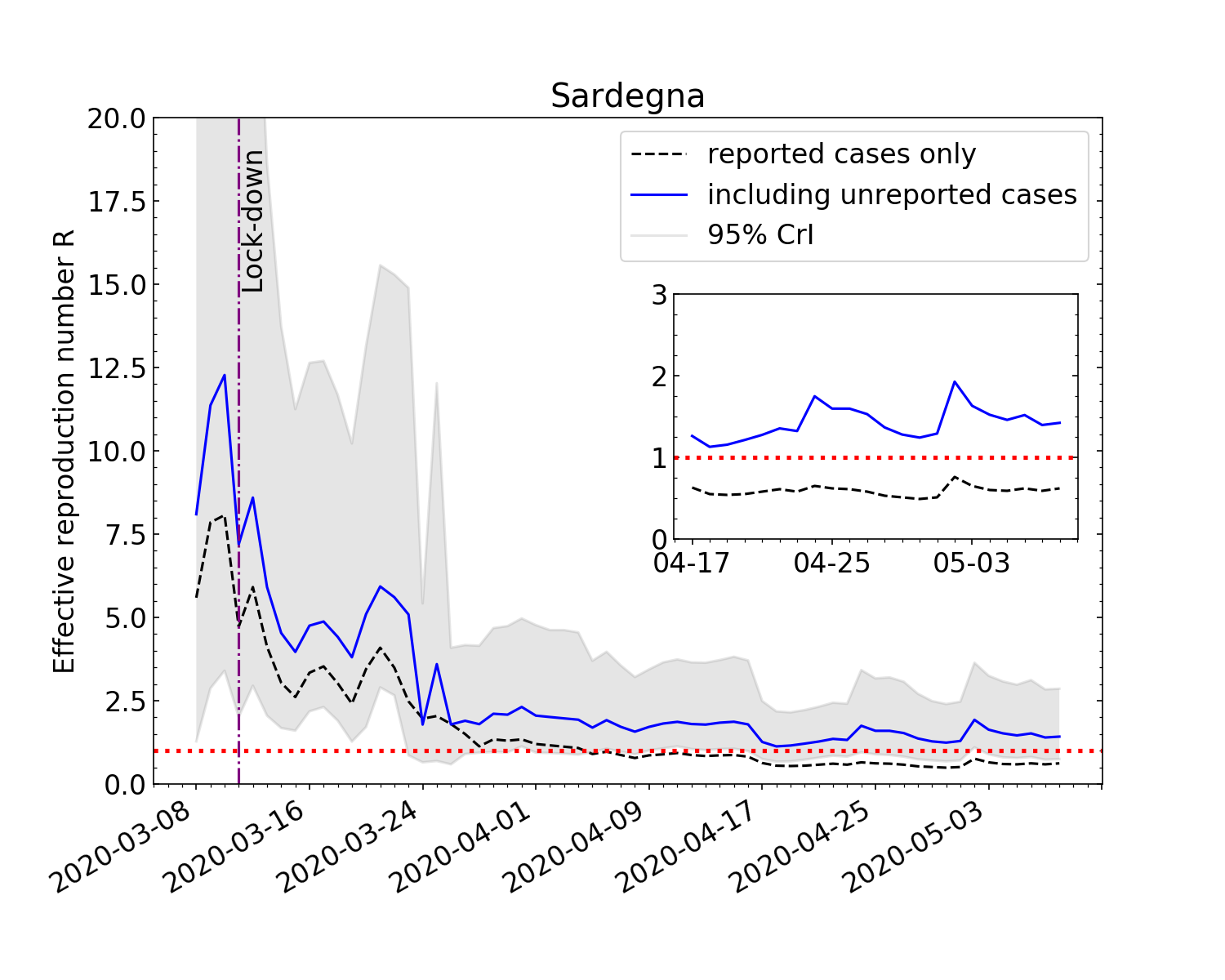}\\
    \includegraphics[width=0.5\textwidth]{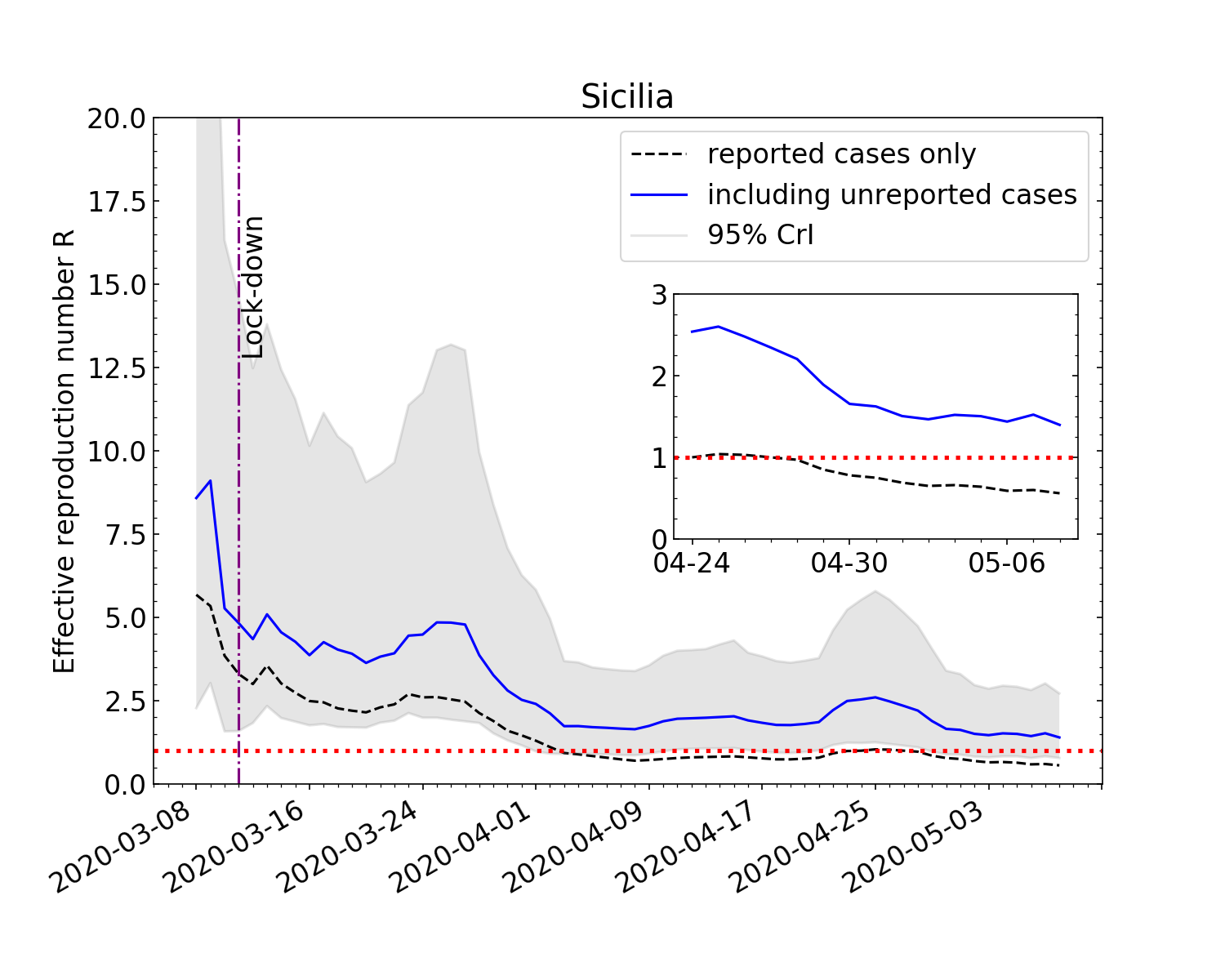}
    \hspace{-0.5cm}
    \includegraphics[width=0.5\textwidth]{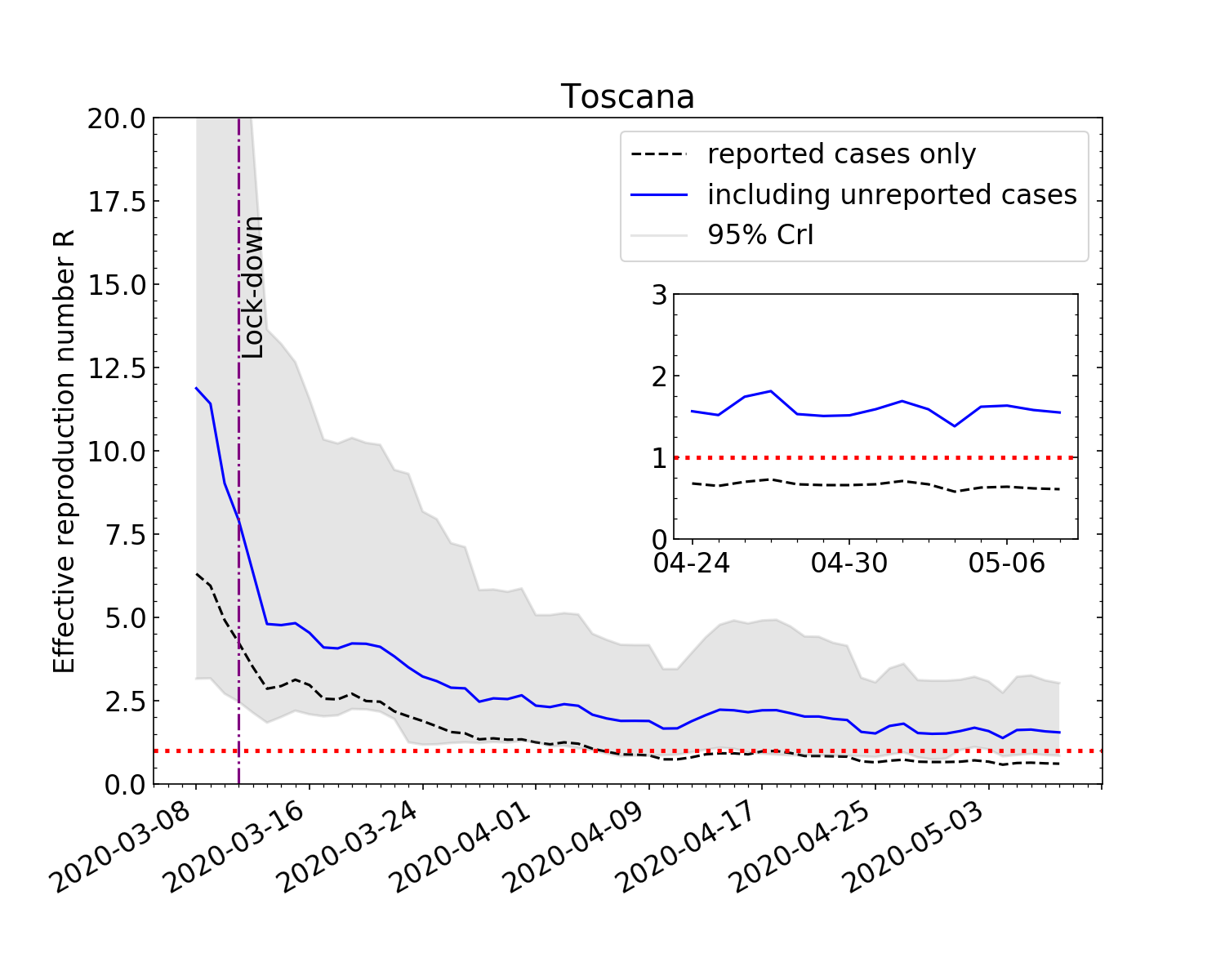}\\
\caption{{\small \emph{Evolution of effective reproduction number for COVID-19
in four Italian regions: Puglia, Sardegna, Sicilia, Toscana.}}}
\label{fig:regions4}
\end{figure}

\begin{figure}[t] 
    \centering
    \includegraphics[width=0.5\textwidth]{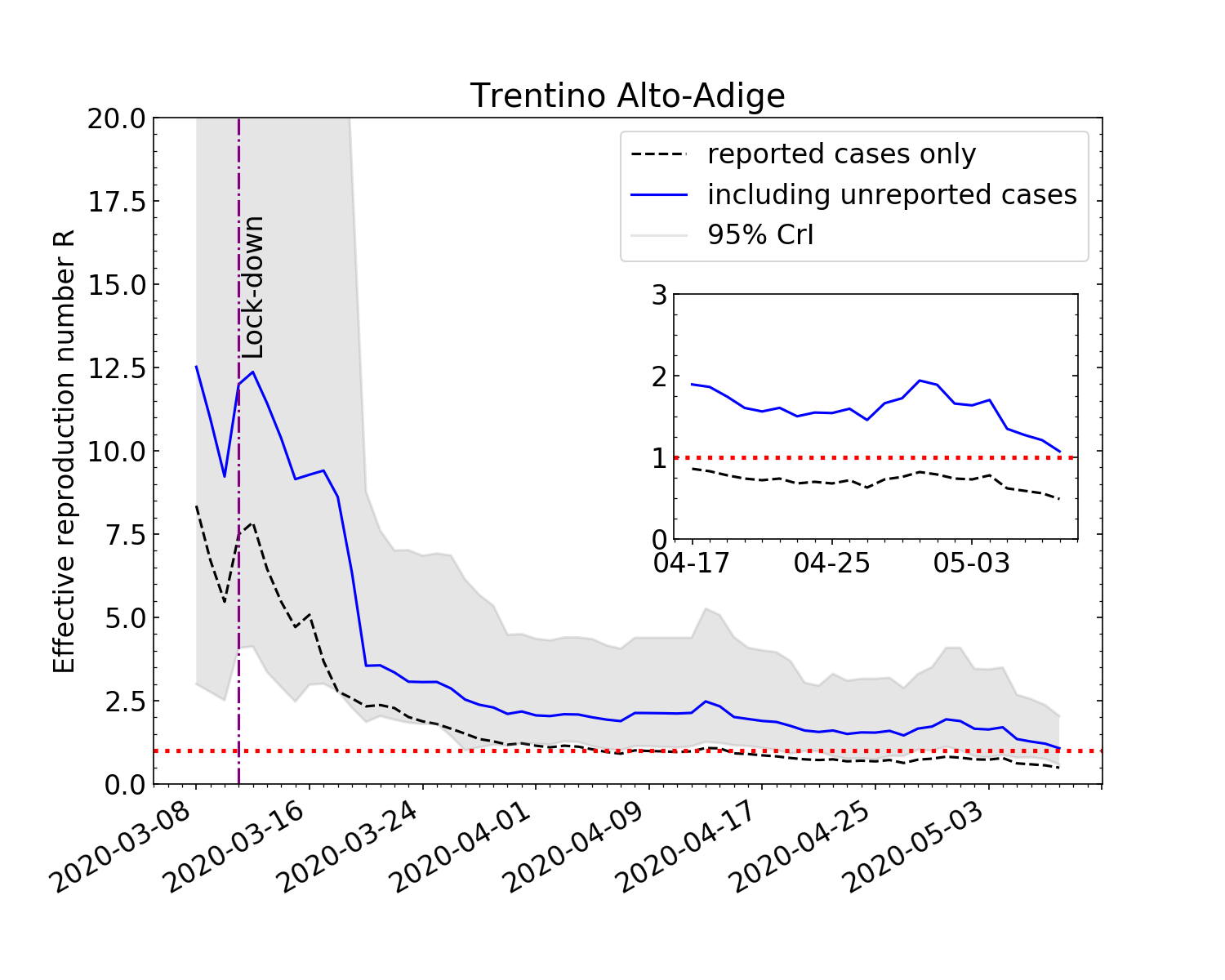}
    \hspace{-0.5cm}
    \includegraphics[width=0.5\textwidth]{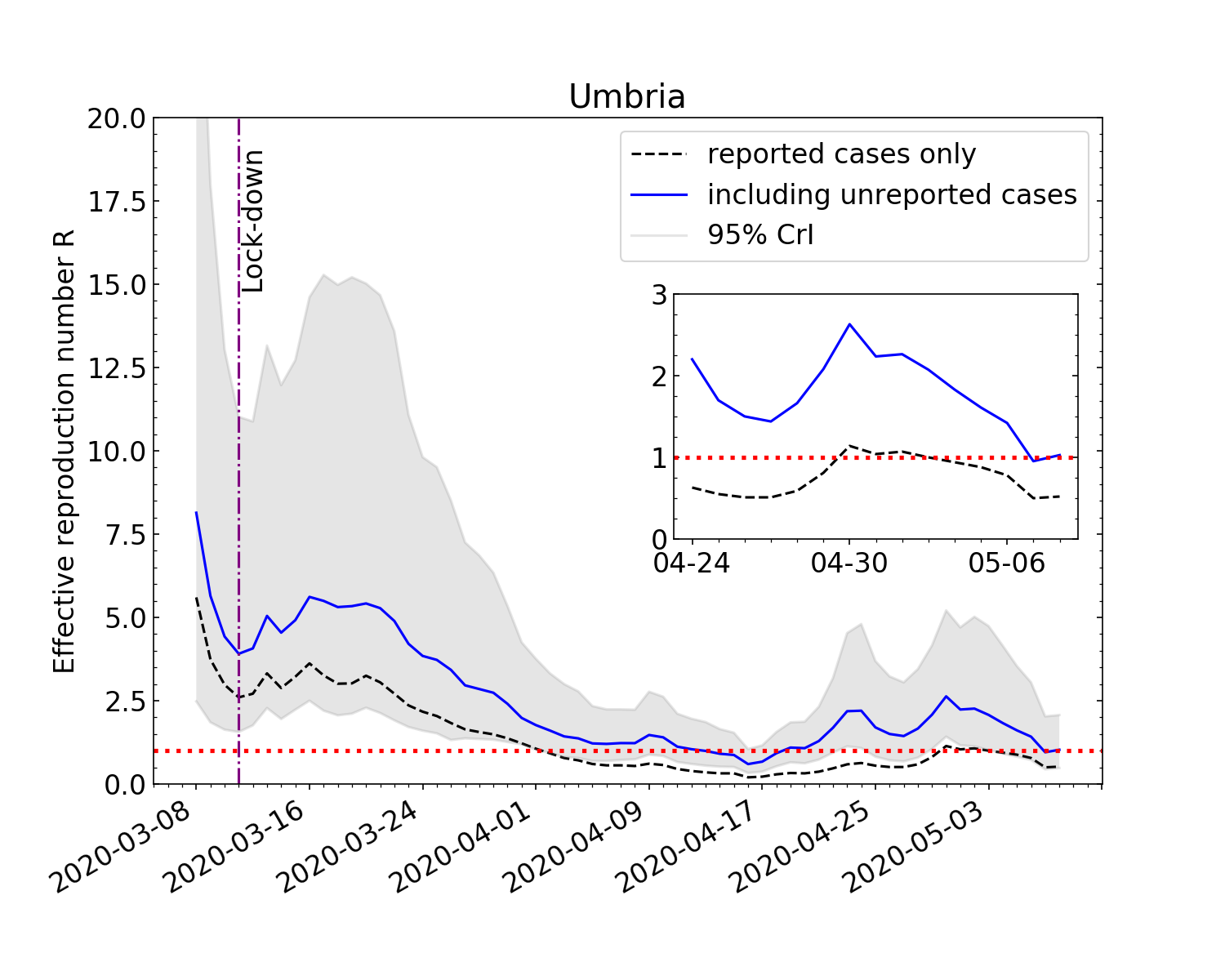}\\
    \includegraphics[width=0.5\textwidth]{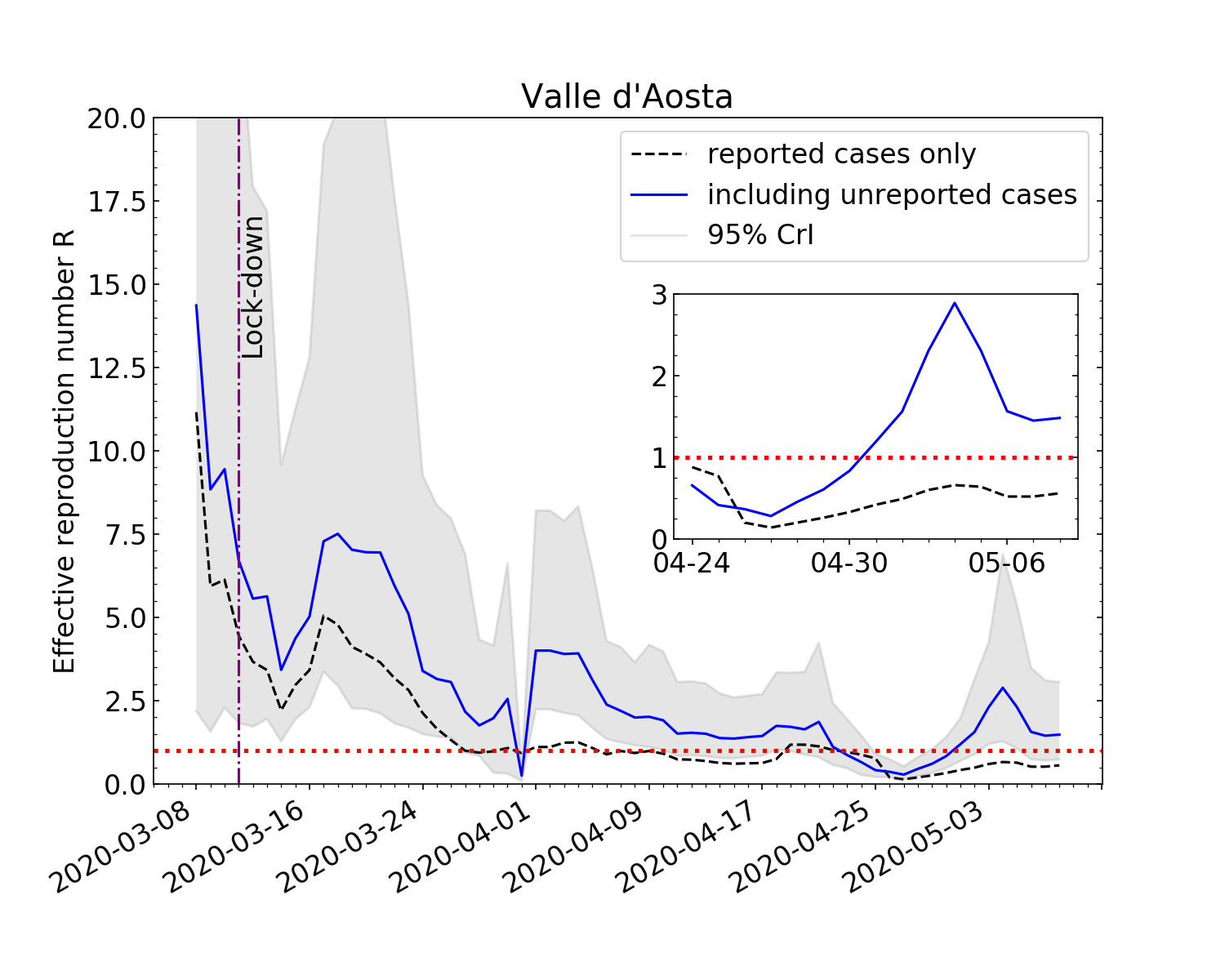}
    \hspace{-0.5cm}
    \includegraphics[width=0.5\textwidth]{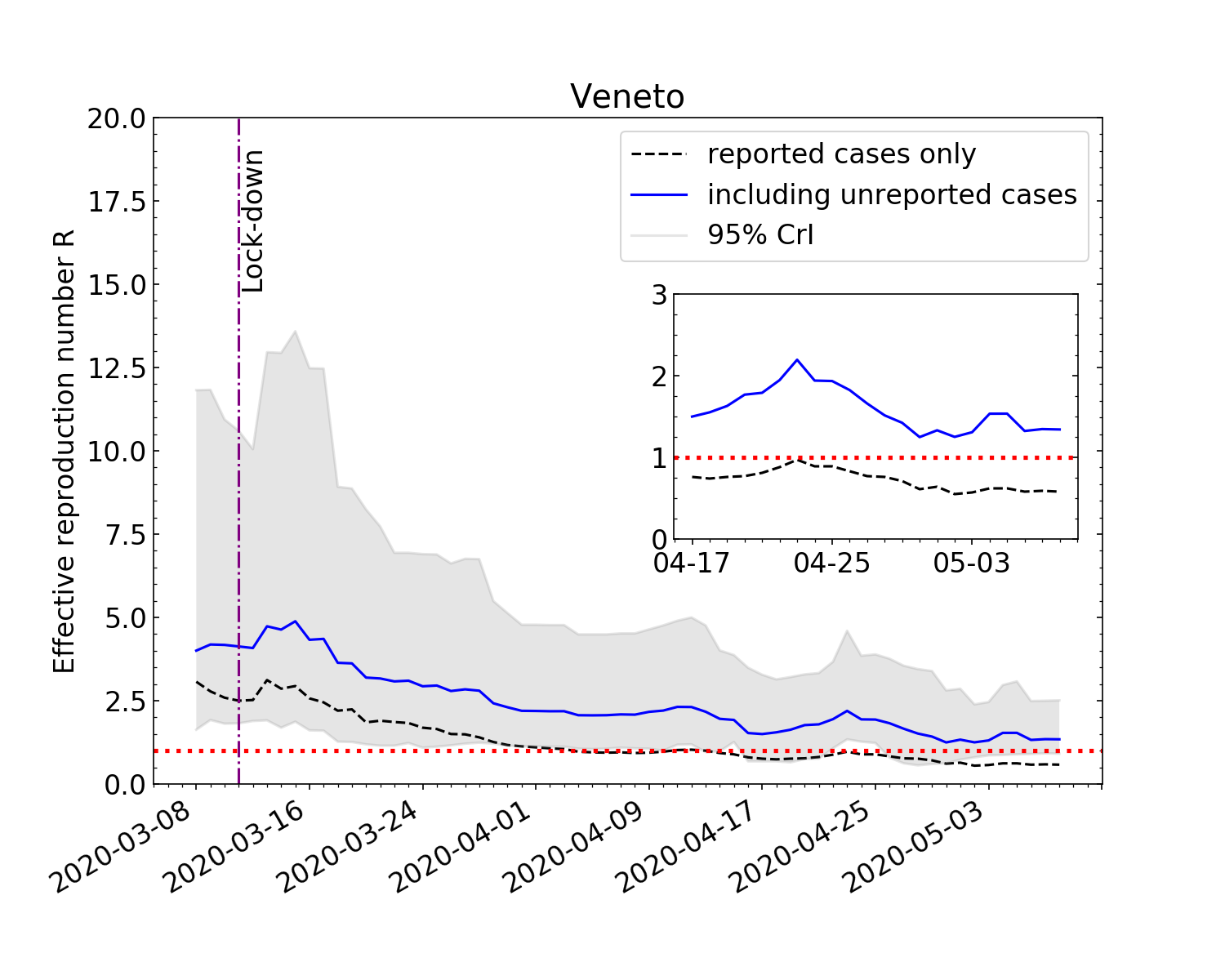}\\
\caption{{\small \emph{Evolution of effective reproduction number  for COVID-19
in four Italian regions: Trentino Alto-Adige, Umbria, Valle d'Aosta, Veneto.}}}
\label{fig:regions5}
\end{figure}

We apply our statistical model, described in 
Section \ref{app:statmodel},  to the incidence data 
in the twenty Italian regions, 
retrieved from Ref.~\cite{protezione_civile}.
The computation of $R_t$ starts 5 days after the first reported day, that is the same for all the regions. This choice is adopted in order to bypass the uncertainty and delay in collected data in the very first days.
For the time window length we use $\tau=7$ days.
The results for all regions in alphabetical order are shown  in Figures \ref{fig:regions1}, \ref{fig:regions2}, \ref{fig:regions3}, 
\ref{fig:regions4}, \ref{fig:regions5}, 
and Table \ref{tab:results_regions}

\newpage

\begin{table}[b]
\centering
\begin{tabular}{|c|c|c|c|}
\hline
\multirow{2}{*}{\textbf{Region}} & 
$R$ \textbf{with reported} &
\textbf{Mean of  marginalized  } &
\multirow{2}{*}{\textbf{95\% CrI}} \\
& \textbf{cases only} & 
\textbf{posterior probability} & \\
\hline
Abruzzo & 0.77 & 2.01 & (1.20,\, 4.03)\\
\hline
Basilicata & 1.29 & 2.94 & (1.24,\, 7.31)\\
\hline
Calabria & 0.46 & 1.32 & (0.64,\, 2.59)\\
\hline
Campania & 0.60 & 1.66 & (0.89,\, 3.36)\\
\hline
{\small{Emila Romagna}} & 0.64 & 1.45 & (0.81,\, 2.93)\\
\hline
Friuli V.~G.& 0.54 & 1.28 & (0.74,\, 2.37)\\
\hline
Lazio & 0.80 & 1.81 & (0.99,\, 3.69)\\
\hline
Liguria & 0.75 & 1.52 & (0.80,\, 3.13)\\
\hline
Lombardia & 0.83 & 2.01 & (0.94,\, 4.14)\\
\hline
Marche & 0.78 & 1.86 & (1.06,\, 3.38)\\
\hline
Molise & 0.65 & 1.83 & (0.72,\, 3.73)\\
\hline
Piemonte & 0.63 & 1.46 & (0.64,\, 3.12)\\
\hline
Puglia & 0.64 & 1.68 & (0.92,\, 3.37)\\
\hline
Sardegna & 0.62 & 1.42 & (0.75,\, 2.86)\\
\hline
Sicilia & 0.56 & 1.39 & (0.78,\, 2.72)\\
\hline
Toscana & 0.61 & 1.55 & (0.85,\, 3.03)\\
\hline
Trentino A.~A. & 0.49 & 1.07 & (0.58,\, 2.04)\\
\hline
Umbria & 0.52 & 1.03 & (0.48,\, 2.07)\\
\hline
Valle d'Aosta & 0.56 & 1.48 & (0.74,\, 3.06)\\
\hline
Veneto & 0.58 & 1.34 & (0.91,\, 2.51)\\
\hline
\end{tabular}
\caption{{\small \emph{
The values of the effective reproduction number $R$  for COVID-19, on  the last day of our analysis (2020-05-08), for each of the Italian regions we considered.
The second column reports the value we find by including only reported incidence data and mean values of the serial interval distribution. On the third column we report the mean of the posterior distribution of $R$, marginalized over the nuisance parameters describing the serial interval and undetected cases distributions.
The corresponding 95\% credible interval is reported on the last column.}}
}
\label{tab:results_regions}
\end{table}

\end{appendices}
\end{document}